%% file: main.tex
\pdfoutput=1
\documentclass[12pt,a4paper]{article}

\usepackage{ifthen} 
\newboolean{pdflatex}
\setboolean{pdflatex}{true} 
\usepackage{tikzscale}
\usepackage[compat=1.1.0]{tikz-feynman}
\usepackage{setspace}
\usepackage{caption,subcaption}
\newboolean{articletitles}
\setboolean{articletitles}{true} 

\newboolean{uprightparticles}
\setboolean{uprightparticles}{false} 

\def\paperauthors{LHCb collaboration} 
\def\paperasciititle{First observation of the Lambdab0 to Lambdac+ Ds- K+ K- decay and search for pentaquarks in the Lambdac+ Ds- system} 
\def\papertitle{First observation of the \texorpdfstring{\decay{\Lb}{\Lc \Dsm \Kp \Km} decay}{Lambda_b0 -> Lambda_c+ D_s- K+ K-} and search for pentaquarks in the \texorpdfstring{\Lc\Dsm}{Lambda_c+ D_s-} system} 
\def\paperkeywords{{High Energy Physics}, {LHCb}} 
\def\papercopyright{\the\year\ CERN for the benefit of the LHCb collaboration} 
\def\paperlicence{CC BY 4.0 licence}
\def\paperlicenceurl{https://creativecommons.org/licenses/by/4.0/}

\newif\ifEnableSectionTOCLinks
\EnableSectionTOCLinksfalse 

\input{preamble}

\providecommand{\LbLcDs}{\decay{\Lb}{\Lc\Dsm}}
\providecommand{\LbLcDsKK}{\decay{\Lb}{\Lc\Dsm\Kp\Km}}
\providecommand{\LbLcDsphi}{\decay{\Lb}{\Lc\Dsm\phi}}
\providecommand{\LbLcDsPhi}{\LbLcDsphi}

\providecommand{\LcpKpi}{\decay{\Lc}{\proton\Km\pip}}
\providecommand{\DsKKpi}{\decay{\Dsm}{\Kp\Km\pim}}

\newcommand{\RBRresultNoBrErr}[3]{\ensuremath{#1\pm#2\pm#3}}

\newcommand{\MyRBFNominal}{\ensuremath{0.0141}}
\newcommand{\MyRBFStat}{\ensuremath{0.0019}}  
\newcommand{\MyRBFSyst}{\ensuremath{0.0012}}

\newcommand{\MyRPcsLow}{\ensuremath{0.12}}
\newcommand{\MyRPcsHigh}{\ensuremath{0.20}}

\newcommand{\RBF}[2]{\ensuremath{\frac{\BF\left(#1\right)}{\BF\left(#2\right)}}\xspace}

\providecommand{\Pcs}{{\ensuremath{P_{\cquark\cquarkbar\squark}}}\xspace}

\providecommand{\Pcsz}{{\ensuremath{P_{\cquark\cquarkbar\squark}^0}}\xspace}

\usepackage{longtable} 

\begin{document}

\renewcommand{\thefootnote}{\fnsymbol{footnote}}
\setcounter{footnote}{1}

\input{title-LHCb-PAPER}


\renewcommand{\thefootnote}{\arabic{footnote}}
\setcounter{footnote}{0}


\cleardoublepage


\pagestyle{plain} 
\setcounter{page}{1}
\pagenumbering{arabic}


\input{body}


\input{acknowledgements}

\section*{Data availability}
The data that support the findings of this article are openly available~\cite{LHCb-PAPER-2025-022-cds}.

\input{appendix}


\addcontentsline{toc}{section}{References}
\bibliographystyle{LHCb}
\bibliography{main,standard,LHCb-PAPER,LHCb-CONF,LHCb-DP,LHCb-TDR}

\newpage
\input{Authorship_LHCb-PAPER-2025-022}

\end{document}

%% file: preamble.tex
\usepackage[top=1in, bottom=1.25in, left=1in, right=1in]{geometry}

\usepackage{microtype}
\usepackage{lineno}  
\usepackage{xspace} 
\usepackage{caption} 

\usepackage{graphicx}  
\usepackage{color}
\usepackage{colortbl}
\graphicspath{{./figs/}} 

\usepackage{amsmath} 
\usepackage{amssymb}
\usepackage{amsfonts}
\usepackage{upgreek} 

\newcommand*\patchAmsMathEnvironmentForLineno[1]{%
\expandafter\let\csname old#1\expandafter\endcsname\csname #1\endcsname
\expandafter\let\csname oldend#1\expandafter\endcsname\csname
end#1\endcsname
 \renewenvironment{#1}%
   {\linenomath\csname old#1\endcsname}%
   {\csname oldend#1\endcsname\endlinenomath}%
}
\newcommand*\patchBothAmsMathEnvironmentsForLineno[1]{%
  \patchAmsMathEnvironmentForLineno{#1}%
  \patchAmsMathEnvironmentForLineno{#1*}%
}
\AtBeginDocument{%
\patchBothAmsMathEnvironmentsForLineno{equation}%
\patchBothAmsMathEnvironmentsForLineno{align}%
\patchBothAmsMathEnvironmentsForLineno{flalign}%
\patchBothAmsMathEnvironmentsForLineno{alignat}%
\patchBothAmsMathEnvironmentsForLineno{gather}%
\patchBothAmsMathEnvironmentsForLineno{multline}%
\patchBothAmsMathEnvironmentsForLineno{eqnarray}%
}

\usepackage{hyperxmp}

\usepackage[pdftex,
            pdfauthor={\paperauthors},
            pdftitle={\paperasciititle},
            pdfkeywords={\paperkeywords},
            pdfcopyright={Copyright (C) \papercopyright},
            pdflicenseurl={\paperlicenceurl}]{hyperref}

\usepackage[colorinlistoftodos,textsize=scriptsize]{todonotes}

\usepackage[bottom,flushmargin,hang,multiple]{footmisc}

\usepackage[all]{hypcap} 

\input{lhcb-symbols-def} 

\hypersetup{
  colorlinks   = true, 
  urlcolor     = blue, 
  linkcolor    = blue, 
  citecolor    = red   
}

\ifEnableSectionTOCLinks
    \usepackage[explicit]{titlesec} 
    
    \let\oldcontentsline\contentsline
    \renewcommand

    \titleformat{\section}{\normalfont\Large\bf}{\hyperlink{tocsection.\thesection}{{\thesection} \parbox[t]{\dimexpr\textwidth-1pc}{#1}}}{1pc}{}

    \titleformat{\subsection}{\normalfont\bf}{\hyperlink{tocsubsection.\thesubsection}{{\thesubsection} \parbox[t]{\dimexpr\textwidth-1pc}{#1}}}{1pc}{}

    \titleformat{name=\section,numberless}[display]{}{}{0pt}{\normalfont\Huge\bfseries #1}
\fi

\usepackage{cite} 
\usepackage{mciteplus}

%% file: lhcb-symbols-def.tex
\usepackage{xspace} 
\usepackage{upgreek}

\def\lhcb   {\mbox{LHCb}\xspace}

\def\MagUp {\mbox{\em Mag\kern -0.05em Up}\xspace}

\ifthenelse{\boolean{uprightparticles}}%
{

 \def\Ppi         {\ensuremath{\uppi}\xspace}

 \def\Pphi        {\ensuremath{\upphi}\xspace}

 \def\Ppsi        {\ensuremath{\uppsi}\xspace}

 \def\PDelta      {\ensuremath{\Delta}\xspace}                 
 \def\PXi         {\ensuremath{\Xi}\xspace}                 
 \def\PLambda     {\ensuremath{\Lambda}\xspace}                 
 \def\PSigma      {\ensuremath{\Sigma}\xspace}                 
 \def\POmega      {\ensuremath{\Omega}\xspace}                 
 \def\PUpsilon    {\ensuremath{\Upsilon}\xspace}
 \let\oldPi\Pi
 \def\PPi         {\ensuremath{\oldPi}\xspace}

 \def\PB      {\ensuremath{\mathrm{B}}\xspace}                 
 \def\PD      {\ensuremath{\mathrm{D}}\xspace}                 
 \def\PJ      {\ensuremath{\mathrm{J}}\xspace}                 
 \def\PK      {\ensuremath{\mathrm{K}}\xspace}                 
 \def\Pb      {\ensuremath{\mathrm{b}}\xspace}                 
 \def\Pc      {\ensuremath{\mathrm{c}}\xspace}

 \def\Pp      {\ensuremath{\mathrm{p}}\xspace}                 

 \def\Ps      {\ensuremath{\mathrm{s}}\xspace}

 \def\thebaroffset{0.0em}
}
{

 \def\Ppi         {\ensuremath{\pi}\xspace}

 \def\Pphi        {\ensuremath{\phi}\xspace}

 \def\Ppsi        {\ensuremath{\psi}\xspace}                 
                  
 \mathchardef\PDelta="7101
 \mathchardef\PXi="7104
 \mathchardef\PLambda="7103
 \mathchardef\PSigma="7106
 \mathchardef\POmega="710A
 \mathchardef\PUpsilon="7107
 \mathchardef\PPi="7105
 \def\PB      {\ensuremath{B}\xspace}                 
 \def\PD      {\ensuremath{D}\xspace}                 
 \def\PJ      {\ensuremath{J}\xspace}                 
 \def\PK      {\ensuremath{K}\xspace}                 
 \def\Pb      {\ensuremath{b}\xspace}                 
 \def\Pc      {\ensuremath{c}\xspace}

 \def\Pp      {\ensuremath{p}\xspace}                 

 \def\Ps      {\ensuremath{s}\xspace}

 \def\thebaroffset{0.18em}
}
\newcommand{\offsetoverline}[2][\thebaroffset]{\kern #1\overline{\kern -#1 #2}}%

\makeatletter
\ifcase \@ptsize \relax
  \newcommand{\miniscule}{\@setfontsize\miniscule{4}{5}}
\or
  \newcommand{\miniscule}{\@setfontsize\miniscule{5}{6}}
\or
  \newcommand{\miniscule}{\@setfontsize\miniscule{5}{6}}
\fi
\makeatother

\DeclareRobustCommand{\optbar}[1]{\shortstack{{\miniscule (\rule[.5ex]{1.25em}{.18mm})}
  \\ [-.7ex] $#1$}}

\def\squark    {{\ensuremath{\Ps}}\xspace}

\def\cquark    {{\ensuremath{\Pc}}\xspace}
\def\cquarkbar {{\ensuremath{\overline \cquark}}\xspace}
\def\ccbar     {{\ensuremath{\cquark\cquarkbar}}\xspace}
\def\bquark    {{\ensuremath{\Pb}}\xspace}

\def\pion   {{\ensuremath{\Ppi}}\xspace}

\def\pip    {{\ensuremath{\pion^+}}\xspace}
\def\pim    {{\ensuremath{\pion^-}}\xspace}

\def\kaon    {{\ensuremath{\PK}}\xspace}

\def\KorKbar {\kern \thebaroffset\optbar{\kern -\thebaroffset \PK}{}\xspace}

\def\Kp      {{\ensuremath{\kaon^+}}\xspace}
\def\Km      {{\ensuremath{\kaon^-}}\xspace}

\newcommand{\phiz}{\ensuremath{\Pphi}\xspace}

\def\Dbar    {{\ensuremath{\offsetoverline{\PD}}}\xspace}
\def\D       {{\ensuremath{\PD}}\xspace}

\def\DorDbar {\kern \thebaroffset\optbar{\kern -\thebaroffset \PD}\xspace}

\def\Dzb     {{\ensuremath{\Dbar{}^0}}\xspace}
\def\Dp      {{\ensuremath{\D^+}}\xspace}
\def\Dm      {{\ensuremath{\D^-}}\xspace}

\def\DpDm    {\ensuremath{\Dp {\kern -0.16em \Dm}}\xspace}

\def\Ds      {{\ensuremath{\D^+_\squark}}\xspace}

\def\Dsm     {{\ensuremath{\D^-_\squark}}\xspace}

\def\B       {{\ensuremath{\PB}}\xspace}

\def\BorBbar {\kern \thebaroffset\optbar{\kern -\thebaroffset \PB}\xspace}

\def\Bd      {{\ensuremath{\B^0}}\xspace}

\def\BdorBdbar {\kern \thebaroffset\optbar{\kern -\thebaroffset \Bd}\xspace}

\def\Bub     {{\ensuremath{\B^-}}\xspace}

\def\Bm      {{\ensuremath{\Bub}}\xspace}

\def\Bs      {{\ensuremath{\B^0_\squark}}\xspace}

\def\BsorBsbar {\kern \thebaroffset\optbar{\kern -\thebaroffset \Bs}\xspace}

\def\jpsi     {{\ensuremath{{\PJ\mskip -3mu/\mskip -2mu\Ppsi}}}\xspace}

\def\Y#1S{\ensuremath{\PUpsilon{(#1S)}}\xspace}

\def\proton      {{\ensuremath{\Pp}}\xspace}
\def\antiproton  {{\ensuremath{\overline \proton}}\xspace}

\def\Lz          {{\ensuremath{\PLambda}}\xspace}

\def\LorLbar     {\kern \thebaroffset\optbar{\kern -\thebaroffset \PLambda}\xspace}

\def\Sigmares    {{\ensuremath{\PSigma}}\xspace}

\def\Xires       {{\ensuremath{\PXi}}\xspace}

\def\Lc          {{\ensuremath{\Lz^+_\cquark}}\xspace}

\def\Sigmac       {{\ensuremath{\Sigmares_\cquark}}\xspace}

\def\Lb           {{\ensuremath{\Lz^0_\bquark}}\xspace}

\def\Xibm         {{\ensuremath{\Xires^-_\bquark}}\xspace}

\def\BF         {{\ensuremath{\mathcal{B}}}\xspace}

\newcommand{\decay}[2]{\mbox{\ensuremath{#1\!\to #2}}\xspace} 

\def\to                 {\ensuremath{\rightarrow}\xspace}

\def\AT#1     {\ensuremath{A_{\mathrm{T}}^{#1}}\xspace}           

\def\C#1      {\ensuremath{\mathcal{C}_{#1}}\xspace}                       
\def\Cp#1     {\ensuremath{\mathcal{C}_{#1}^{'}}\xspace}                    
\def\Ceff#1   {\ensuremath{\mathcal{C}_{#1}^{\mathrm{(eff)}}}\xspace}        
\def\Cpeff#1  {\ensuremath{\mathcal{C}_{#1}^{'\mathrm{(eff)}}}\xspace}       
\def\Ope#1    {\ensuremath{\mathcal{O}_{#1}}\xspace}                       
\def\Opep#1   {\ensuremath{\mathcal{O}_{#1}^{'}}\xspace}                    

\newcommand{\aunit}[1]{\ensuremath{\text{\,#1}}}       

\newcommand{\tev}{\aunit{Te\kern -0.1em V}\xspace}
\newcommand{\gev}{\aunit{Ge\kern -0.1em V}\xspace}
\newcommand{\mev}{\aunit{Me\kern -0.1em V}\xspace}
\newcommand{\kev}{\aunit{ke\kern -0.1em V}\xspace}
\newcommand{\ev}{\aunit{e\kern -0.1em V}\xspace}
 
\newcommand{\mevc}{\ensuremath{\aunit{Me\kern -0.1em V\!/}c}\xspace}
\newcommand{\gevc}{\ensuremath{\aunit{Ge\kern -0.1em V\!/}c}\xspace}
\newcommand{\mevcc}{\ensuremath{\aunit{Me\kern -0.1em V\!/}c^2}\xspace}
\newcommand{\gevcc}{\ensuremath{\aunit{Ge\kern -0.1em V\!/}c^2}\xspace}

\def\fb   {\ensuremath{\aunit{fb}}\xspace}
\def\invfb   {\ensuremath{\fb^{-1}}\xspace}

\newcommand{\chisq}{\ensuremath{\chi^2}\xspace}

\newcommand{\chisqip}{\ensuremath{\chi^2_{\text{IP}}}\xspace}

\newcommand{\chisqvtx}{\ensuremath{\chi^2_{\text{vtx}}}\xspace}

\def\gsim{{~\raise.15em\hbox{$>$}\kern-.85em
          \lower.35em\hbox{$\sim$}~}\xspace}
\def\lsim{{~\raise.15em\hbox{$<$}\kern-.85em
          \lower.35em\hbox{$\sim$}~}\xspace}

\def\sPlot{\mbox{\em sPlot}\xspace}

\def\pt         {\ensuremath{p_{\mathrm{T}}}\xspace}

\def\mrad{\aunit{mrad}\xspace}

\def\evtgen     {\mbox{\textsc{EvtGen}}\xspace}

\def\geant      {\mbox{\textsc{Geant4}}\xspace}

\def\photos     {\mbox{\textsc{Photos}}\xspace}

\def\pythia     {\mbox{\textsc{Pythia}}\xspace}

\def\tell1  {TELL1\xspace}
\def\ukl1   {UKL1\xspace}

\newcommand{\lhcborcid}[1]{\href{https://orcid.org/#1}{\hspace*{0.1em}\raisebox{-0.45ex}{\includegraphics[width=1em]{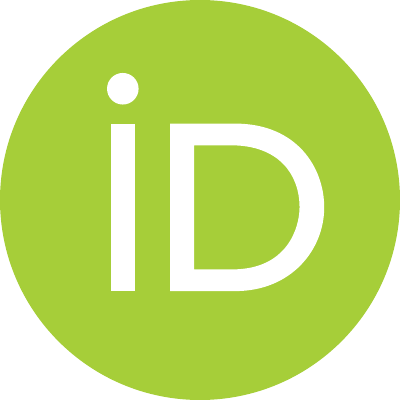}}}}


%% file: title-LHCb-PAPER.tex

\begin{titlepage}
\pagenumbering{roman}

\vspace*{-1.5cm}
\centerline{\large EUROPEAN ORGANIZATION FOR NUCLEAR RESEARCH (CERN)}
\vspace*{1.5cm}
\noindent
\begin{tabular*}{\linewidth}{lc@{\extracolsep{\fill}}r@{\extracolsep{0pt}}}
\vspace*{-1.5cm}\mbox{\!\!\!\includegraphics[width=.14\textwidth]{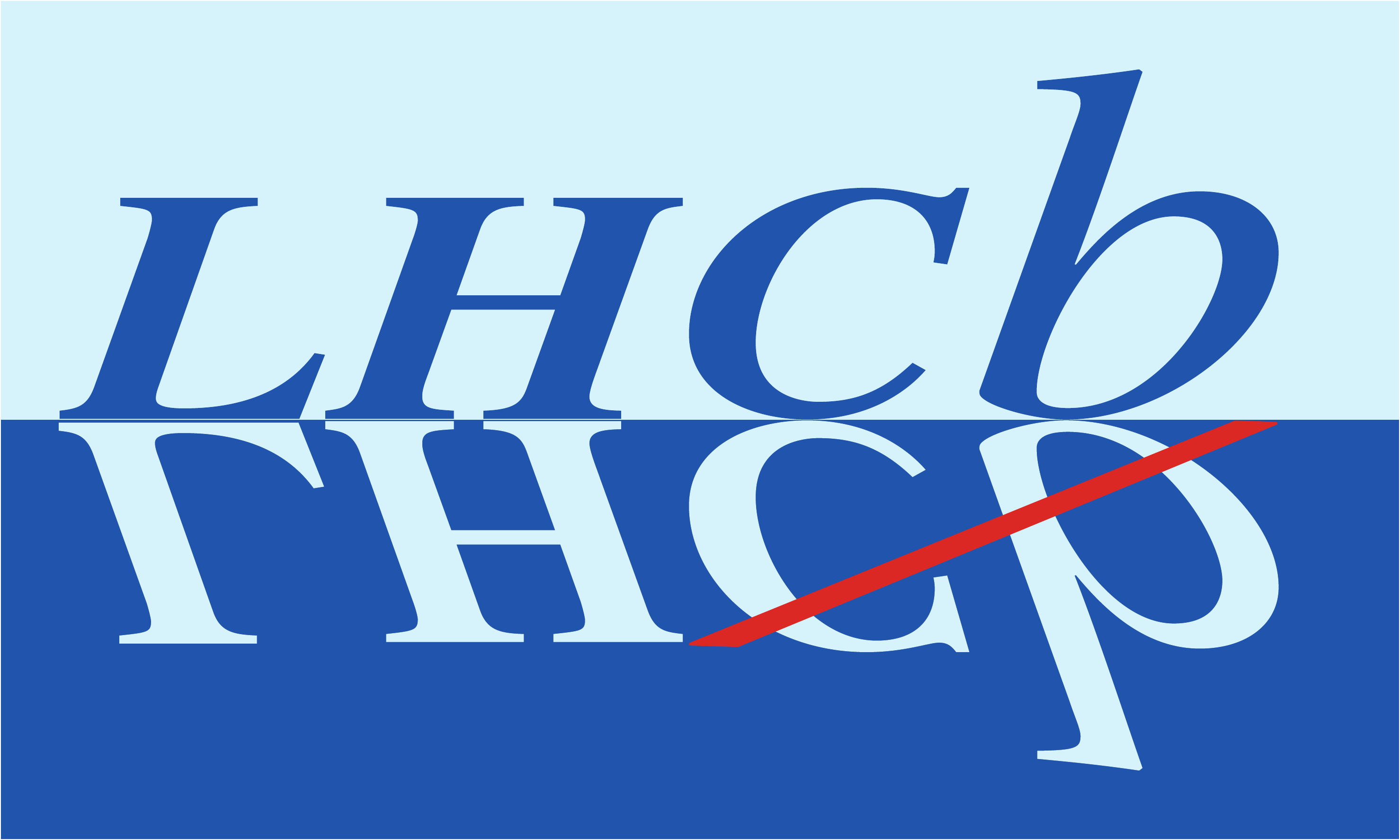}} & &%
\\
 & & CERN-EP-2025-138 \\  
 & & LHCb-PAPER-2025-022 \\  
 & & September 29, 2025 \\ 
 & & \\
\end{tabular*}

\vspace*{4.0cm}

{\normalfont\bfseries\boldmath\huge
\begin{center}
  \papertitle 
\end{center}
}

\vspace*{2.0cm}

\begin{center}
\paperauthors\footnote{Full author list given at the end of the article.}
\end{center}

\vspace{\fill}

\begin{abstract}
  \noindent
  The \mbox{$\mathit{\Lambda}_b^{0}\!\rightarrow\mathit{\Lambda}_{c}^{+}D_{s}^{-}K^{+}K^{-}$} decay is observed for the first time using the data sample from proton-proton collisions recorded at a center-of-mass energy of $13\,\aunit{Te\kern -0.1em V}$ with the LHCb detector, corresponding to an integrated luminosity of $6\,\aunit{fb}^{-1}$. The ratio of branching fraction to that of \mbox{$\mathit{\Lambda}_b^{0} \!\rightarrow\mathit{\Lambda}_{c}^{+}D_{s}^{-}$} decays is measured as \mbox{$0.0141 \pm 0.0019 \pm 0.0012$}, where the first uncertainty is statistical and the second systematic. A search for hidden-charm pentaquarks with strangeness is performed in the \mbox{$\mathit{\Lambda}_{c}^{+}D_{s}^{-}$} system. No evidence is found, and upper limits on the production ratio of $P_{c\bar{c}s}(4338)^0$ and $P_{c\bar{c}s}(4459)^0$ pentaquarks relative to the \mbox{$\mathit{\Lambda}_{c}^{+}D_{s}^{-}$} final state are set at the $95\%$ confidence level as $0.12$ and $0.20$, respectively.
\end{abstract}

\vspace*{2.0cm}

\begin{center}
  Published in Phys.~Rev.~D112 (2025) 052013
\end{center}

\vspace{\fill}

{\footnotesize 
\centerline{\copyright~\papercopyright. \href{\paperlicenceurl}{\paperlicence}.}}
\vspace*{2mm}

\end{titlepage}


\newpage
\setcounter{page}{2}
\mbox{~}
%
%
%
%

%% file: body.tex
\section{Introduction}
\label{sec:Introduction}

Pentaquarks with minimal quark composition of three quarks plus a quark-antiquark pair have been predicted since the establishment of the quark model~\cite{GellMann:1964nj,*Zweig:352337}.
Hidden-charm pentaquark candidates ($P_{\ccbar}^+=\ccbar uud$) were first observed in the \mbox{\jpsi\proton} system, produced through \decay{\Lb}{\jpsi\proton\Km} decays recorded by the \lhcb experiment~\cite{LHCb-PAPER-2015-029,LHCb-PAPER-2016-009,LHCb-PAPER-2019-014}.\footnote{Inclusion of charge-conjugate processes is implied throughout.} 
A pentaquark candidate ($P_{\ccbar s}^0=\ccbar uds$) with strangeness was observed in the \mbox{\jpsi\Lz} system in \decay{\Bm}{\jpsi\Lz\antiproton} decays~\cite{LHCb-PAPER-2022-031}. Besides, evidence for candidate $P_{\ccbar}^+$ and $P_{\ccbar s}^0$ resonances was found in the \decay{\Bs}{\jpsi\proton\antiproton} decay and \decay{\Xibm}{\jpsi\Lz\Km} decay, respectively~\cite{LHCb-PAPER-2021-018,LHCb-PAPER-2020-039}.
The $P_{\ccbar}^+$ candidates are found near the threshold for the production of \mbox{$\PSigma_{c}^{+}\smash{\Dbar}^{(*)0}$} states, and $P_{\ccbar\squark}^{0}$ candidates are found close to the threshold of \mbox{$\PXi_{c}\Dbar^{(*)}$} states.
Various interpretations, including hadronic molecules~\cite{Wu:2010jy,Wu:2010vk,Yang:2011wz,Dong:2021juy,Guo:2017jvc}, tightly bound pentaquark states~\cite{Yuan:2012wz,Maiani:2015vwa,Lebed:2015tna,Li:2015gta,Esposito:2016noz,Richard:2016eis,Zhu:2015bba} and threshold effects~\cite{Guo:2015umn,Liu:2015fea,Bayar:2016ftu,Guo:2019twa}, have been proposed, but their nature remains largely unknown. Measurements in different decay channels are crucial for testing and distinguishing these interpretations~\cite{Chen:2016qju,Dong:2017gaw,Olsen:2017bmm}.

Several searches for pentaquarks have been conducted in open-charm final states at \lhcb. 
The prompt production of hidden-charm pentaquark states decaying to \mbox{$\Sigmac\Dbar$} and \mbox{$\Lc\Dbar$}, as well as doubly charmed pentaquark states decaying to \mbox{$\Sigmac D$} and \mbox{$\Lc D$}, have been studied~\cite{LHCB-PAPER-2023-018}.
Additionally, the observation of \decay{\Lb}{\Lc \Dzb \Km} and  \decay{\Lb}{\PSigma_{c}^{(*)++} D^{(*)-}\Km} decays allows for testing pentaquark states in the \Lc\Dzb and \mbox{$\PSigma_{c}^{(*)++} D^{(*)-}$} systems, respectively~\cite{LHCb-PAPER-2023-034,LHCb-paper-2023-044}.
However, no evidence for such states has been found in these studies.

\begin{figure}[tb]
    \centering
    \subfloat[{The nonresonant $\Kp\Km$ process}]{
        \includegraphics[width=0.45\linewidth]{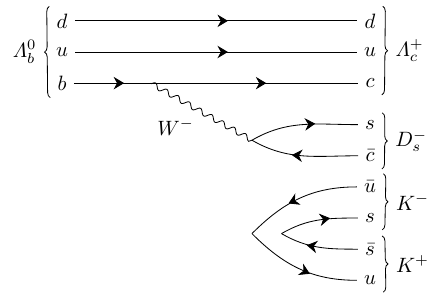}
        \label{fig:Feynman_signalKK}
        }
    \subfloat[{The $\phi$-dominated process}]{
        \includegraphics[width=0.45\linewidth]{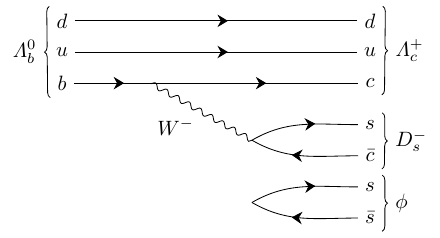}
        \label{fig:Feynman_signalphi}
    }
    \caption{Tree-level Feynman diagrams of the \decay{\Lb}{\Lc \Dsm \Kp \Km} decay.}
    \label{fig:Feynman_signal}
\end{figure}

The \decay{\Lb}{\Lc \Dsm \Kp \Km} decay offers a special opportunity to search for pentaquark candidates in the \mbox{\Lc\Dsm} system, whose leading Feynman diagrams are shown in Fig.~\ref{fig:Feynman_signal}.
In particular, in the \decay{\Lb}{\Lc\Dsm\phi} process, the phase space of the \Lc\Dsm system extends from \mbox{$4235$} to \mbox{$4632\mev$},\footnote{Natural units with $\hbar = c = 1$ are used throughout.} and, thus, encompasses the masses of the two pentaquark candidates $\Pcs(4338)^0$ and $\Pcs(4459)^0$, previously observed in the decays \decay{\Bm}{\jpsi\Lz\antiproton} and \decay{\Xibm}{\jpsi\Lz\Km}, respectively. Meanwhile, searches for highly excited $\Lc$ and $\Dsm$ states, as well as exotic particles in the $\Lc\phi$ and $\Dsm\phi$ systems, can also be performed.

In this paper, the first observation of the \decay{\Lb}{\Lc \Dsm \Kp \Km} decay is reported. The relative branching fraction with respect to the \LbLcDs decay is measured, and a search for pentaquark candidates in the \Lc\Dsm system is performed, using proton-proton ($\proton\proton$) collision data at a center-of-mass energy of 13\tev, corresponding to a luminosity of 6\invfb, collected by the \lhcb experiment between 2015 and 2018. The Cabibbo-favored decay \LbLcDs is chosen as the normalization mode for the branching fraction measurement due to its larger statistics and the same parent particle and intermediate states. It also serves as a control mode for applying corrections to simulated samples using a data-driven method.

\section{Detector and simulation}
The \lhcb detector~\cite{LHCb-DP-2008-001,LHCb-DP-2014-002} is a
single-arm forward spectrometer covering the pseudorapidity range $2 < \eta < 5$, designed for the study of particles containing \bquark\ or \cquark\ quarks. The detector elements that are particularly relevant to this analysis are: a silicon-strip vertex locator (VELO) surrounding the $pp$ interaction region that allows \cquark\ and \bquark\ hadrons to be identified from their characteristically long flight distance; a tracking system that provides a measurement of the momentum, $p$, of charged particles; and two ring-imaging Cherenkov detectors that are able to discriminate between different species of charged hadrons.

The online event selection is performed by a trigger which consists of a hardware stage followed by a two-level software stage~\cite{LHCb-DP-2012-004}. At the hardware stage, events are required to have a muon with high momentum transverse to the beam, \pt, or a hadron, photon, or electron with high transverse energy in the calorimeters. For hadrons, the transverse energy threshold is $3.5\gev$. In the offline selection, the software trigger requires a two-, three-, or four-track secondary vertex with a significant displacement from any primary $pp$ interaction vertex (PV). At least one charged particle must have a $\pt > 1.6\gev$ and be inconsistent with originating from a PV. A multivariate algorithm~\cite{BBDT,LHCb-PROC-2015-018} is used for the identification of secondary vertices consistent with the decay of a \Lb hadron.

Simulation is required to model the effects of the detector acceptance and the imposed selection requirements. In the simulation, $pp$ collisions are generated using \pythia~\cite{Sjostrand:2007gs,Sjostrand:2006za}  with a specific \lhcb configuration~\cite{LHCb-PROC-2010-056}. Decays of unstable particles are described by \evtgen~\cite{Lange:2001uf}, in which final-state radiation is generated using \photos~\cite{davidson2015photos}. The interaction of the generated particles with the detector, and its response, are implemented using the \geant toolkit~\cite{Allison:2006ve, *Agostinelli:2002hh} as described in Ref.~\cite{LHCb-PROC-2011-006}. 

The kinematic distribution of \Lb baryons is not perfectly modeled in the simulation framework. Differences in \pt and $\eta$ distributions between data and simulated events for both signal and normalization modes are corrected using a data-driven approach. Additionally, the simulated decay amplitudes for \LcpKpi and \DsKKpi decays do not fully match those observed in data, and corrections for the \Lc and \Dsm decay amplitudes are derived from \LbLcDs decays by comparing their Dalitz plots in data and simulation.

\section{Dataset and selection}
The data undergo additional filtering in an offline selection process. Candidates consistent with the exclusive decays \decay{\Lb}{\Lc\Dsm\Kp\Km} or \decay{\Lb}{\Lc\Dsm} are retained, where the \Lc baryon decays to \mbox{\proton\Km\pip} and the \Dsm meson decays to \mbox{\Kp\Km\pim}. The selection requires tracks with hits in at least the VELO and the tracking stations downstream of the magnet~\cite{LHCb-DP-2023-004}. A further selection is applied, requiring that each final-state track has $\pt>100\mev$ and good track-fit quality, and that the \Lb baryon has a significant displacement of its decay vertex from any PV. 
Intermediate charmed hadrons are selected using the quality of their decay-vertex fit (\chisqvtx), the displacement of the decay vertex from the associated PV, and the angle between the reconstructed momentum direction and the flight direction determined from the origin and the decay vertices (DIRA). 
Even though the \Lc decay produces protons with higher momentum that could be cleanly reconstructed and identified, the shorter lifetime compared to that of the \Dsm meson complicates the suppression of background contributions in the decay \decay{\Lc}{\proton\Km\pip}. Additional kinematic and topological criteria are employed to address this issue, based on the sum of \pt of \Lc decay products and the distance of closest approach of these decay products. 
Consequently, the \decay{\Lc}{\proton\Km\pip} decay is selected using a gradient-boosted decision tree (BDT) classifier
~\cite{GBDT,Hocker:2007ht} which is trained and validated with \decay{\Lb}{\Lc\pim} data~\cite{LHCb-PAPER-2014-002,LHCb-PAPER-2023-034}.

The \Lb baryon is reconstructed by combining a \Lc baryon with a \Dsm meson for the control mode and together with two additional kaon tracks for the signal mode.
Kinematical and topological selections are applied, including \pt, \chisqvtx, DIRA, $\chisqip$ (defined as the difference in the vertex fit \chisq of a primary vertex reconstructed with and without the \Lb candidate), and the quality of the kinematic fit, which requires the \Lb to originate from its associated primary vertex~\cite{Hulsbergen:2005pu}. The particle identification (PID) probability of each final-state track, derived from a neural network combining PID information across the detector, is used to suppress topologically similar backgrounds~\cite{LHCb-DP-2014-002}. To enhance the mass resolution, the mass of the \Lb baryon for both signal and normalization modes is derived from a global decay chain fit with a Kalman filter~\cite{Hulsbergen:2005pu}, which constrains the \Lb to originate from its associated PV and the \Lc and \Dsm masses to their known values~\cite{PDG2024}.

A dedicated BDT classifier is trained using simulated events as the signal proxy and the upper sideband data  $\left[m(\Lc\Dsm\Kp\Km) \in [5700, 6200] \mev\right]$ as the background proxy, to suppress combinatorial background in \Lb decays. The BDT employs the AdaBoost algorithm~\cite{AdaBoost,Hocker:2007ht} to enhance its performance. Several kinematical variables, PID variables, and the vertex-fit quality are used as discriminating features. The BDT output selection criterion is chosen as in Ref.~\cite{LHCb-PAPER-2022-031} by maximizing the figure of merit \mbox{$S^2/(S+B)^{3/2}$} to obtain both high signal purity and significance, where $S$ and $B$ are the signal and background yields, respectively, in a region of $\pm12\mev$ around the known \Lb mass. To avoid a possible bias due to fluctuations of the signal yield, $S$ is determined from a fit to the \mbox{$m(\Lc\Dsm\Kp\Km)$} distribution in data selected with a looser BDT requirement and then multiplied by the BDT efficiency derived from simulated samples. The determination of $B$ is obtained by fitting the $\Lb$ mass distribution within \mbox{$m(\Lc\Dsm\Kp\Km) \in [5540, 5590] \cup [5700, 5950] \mev$} and subsequently extrapolating linearly into the signal region. An independent BDT classifier is trained to select normalization mode candidates using a similar strategy and is optimized with the same figure of merit.

Candidates with an opening angle between any track pair smaller than $0.5\mrad$ are rejected, removing artifacts from wrong matching of track segments reconstructed within the tracking stations and the VELO.

Some events passing all selection criteria contain multiple candidates. In such cases, a single candidate is retained at random. The potential bias arising from this random selection is subsequently evaluated as a systematic uncertainty.

\section{Signal extraction}
The yield of the \LbLcDsKK signal decay is determined through an unbinned maximum-likelihood fit to the \mbox{$m(\Lc\Dsm\Kp\Km)$} mass distribution. The \Lb signal is modeled using a linear combination of a Crystal Ball distribution~\cite{Skwarnicki:1986xj} and a Johnson $S_U$ distribution~\cite{JohnsonSU}. These distributions share identical peak positions and widths, which are floated in the fit, while the tail parameters and the relative intensity are fixed to values obtained from fits to simulated data. The combinatorial background is described by a first-order Chebyshev polynomial.

\begin{figure}[tb]
    \centering    
    \includegraphics[width=0.5\linewidth]{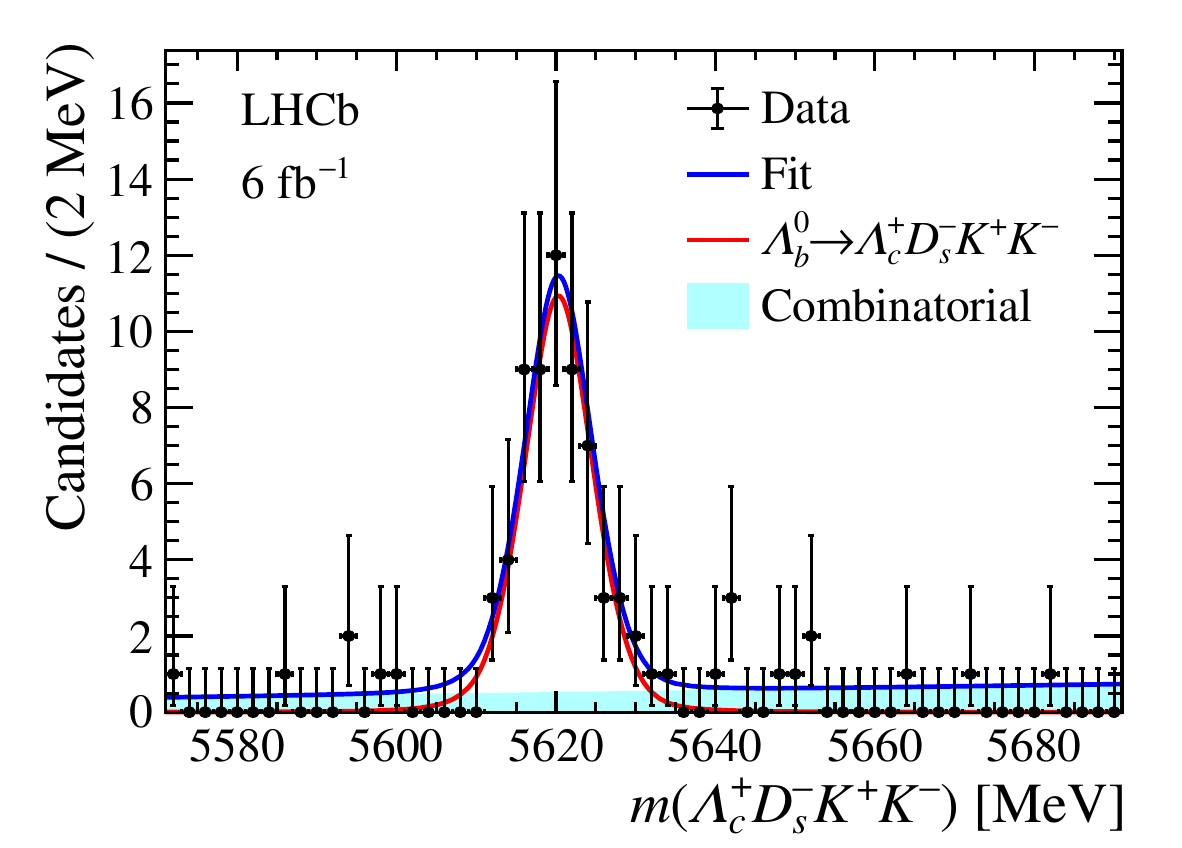}
    \caption{Mass distribution of \Lb candidates reconstructed using the \LbLcDsKK decays, with the fit result also shown.}
    \label{fig:Lb_fit}
\end{figure}

The fit result, shown in Fig.~\ref{fig:Lb_fit}, yields $61\pm 8$ signal 
\LbLcDsKK decays, where the uncertainty is statistical only. The performance of the maximum-likelihood fit is validated by generating and fitting 500 pseudoexperiments using the baseline model, with no significant bias observed for the signal decay mode.

\begin{figure}[tb]
    \centering
    \includegraphics[width=0.5\linewidth]{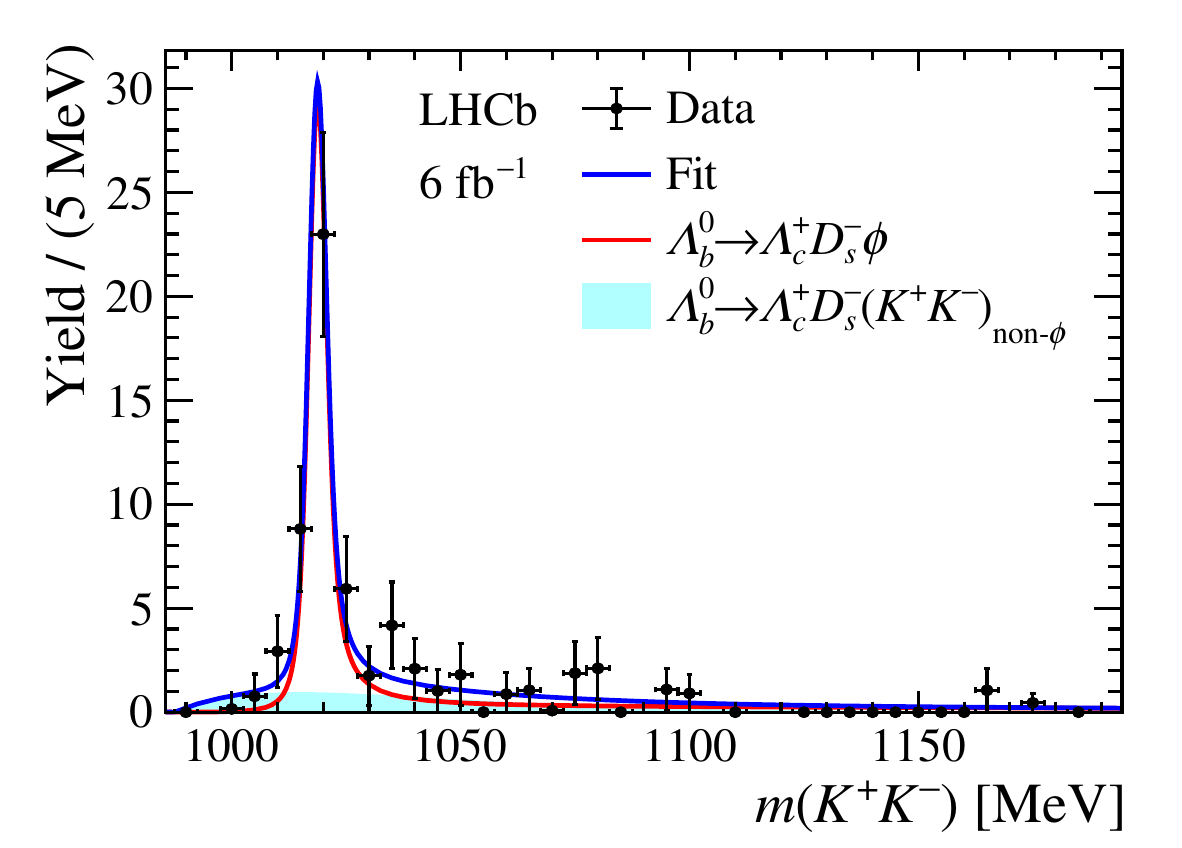}
    \caption{Distribution of \mbox{$m(\Kp\Km)$} for background-subtracted \LbLcDsKK decays with the fit result also shown.}
    \label{fig:KK_mass}
\end{figure}

The contribution from the \LbLcDsPhi process is determined by fitting the background-subtracted $m(\Kp\Km)$  distribution in \LbLcDsKK decays, where backgrounds are subtracted using the \sPlot technique~\cite{Pivk:2004ty}. The \phiz signal is modeled with a relativistic Breit-Wigner distribution, while the remaining non-\phiz component is described by an empirical function. As is illustrated in Fig.~\ref{fig:KK_mass}, the \phiz component is found to be significantly more prominent than the non-\phiz contribution. Furthermore, analysis of the background-subtracted \mbox{$\Lc\phiz$} and \mbox{$\Dsm\phiz$} invariant mass distributions shows no evidence for resonant states.

The yield of the normalization mode \LbLcDs is determined through a fit to the \mbox{$m(\Lc\Dsm)$} distribution. Both the signal and background components for the normalization decay are modeled in the same manner as for the signal decay. The fitted yield for the \LbLcDs decay mode is determined to be \mbox{$(31.27 \pm 0.19)\times 10^3$}, where the uncertainty is statistical only. The same validation procedure used for the signal decay mode is applied to the normalization decay mode, resulting in a relative yield shift of $0.2\%$, which is assigned as a systematic uncertainty.

\section{Relative branching fraction}

In order to determine the ratio of branching fractions, the efficiencies for the signal and normalization channels are estimated from corrected simulated samples, including the effects of detector acceptance, triggers, reconstruction, and offline selection. Since the intermediate states in \LbLcDsKK decays are not perfectly modeled in the simulation, the yield of the signal mode is corrected by considering a per-event efficiency,  
\begin{equation}
    n_\textrm{corr} = \sum_i \frac{s\mathcal{W}_i}{\epsilon(m_i^2(\Lc\Kp\Km), m_i^2(\Dsm\Kp\Km))},
    \label{equ:eff_corr_yield}
\end{equation}
where $n_\textrm{corr}$ is the efficiency-corrected yield, $s\mathcal{W}_i$ is the per-event weight used to subtract background, and \mbox{${\epsilon(m_i^2(\Lc\Kp\Km), m_i^2(\Dsm\Kp\Km))}$} is the mass-dependent per-event efficiency.

The efficiency-corrected yield for the two-body decay \LbLcDs can be simplified as $
    n(\LbLcDs)/\epsilon(\LbLcDs)
    $, where $n$ and $\epsilon$ represent the yield and efficiency of the normalization decay mode, respectively.  Using these quantities, the ratio of branching fractions is calculated as 
\begin{equation}
    \RBF{\LbLcDsKK}{\LbLcDs} = 
    \frac{n_\textrm{corr}\left(\LbLcDsKK\right)}{n_\textrm{corr}\left(\LbLcDs\right)},
    \label{equ:RBF_Calc}
\end{equation}
yielding a value of ${\MyRBFNominal}\pm{\MyRBFStat}$, where the uncertainty is statistical only.
\section{Systematic uncertainties}
The signal decay \LbLcDsKK includes two additional kaon tracks compared to the normalization \LbLcDs decay. The modeling of track reconstruction efficiency in the simulation introduces a systematic uncertainty of $1.36\%$ for each extra kaon, arising from the imperfect simulation of hadronic interactions with the detector material~\cite{LHCb-DP-2013-002}.

Imperfections in the estimation of the hardware trigger efficiencies are present in both signal mode and normalization mode. A data-driven approach~\cite{LHCb-DP-2012-004} is used to calculate the correction factor in bins of the maximum $\pt$ of the final-state tracks. 
In the baseline measurement, independent correction tables derived from data for the corresponding decays are applied to the \LbLcDsKK and \LbLcDs simulations. Systematic uncertainties in these corrections are estimated by either using a single correction table for both decay modes or by applying the correction factors derived from \LbLcDsKK and \LbLcDs decays to the \LbLcDs and \LbLcDsKK simulations, respectively. The largest change in the efficiency ratio between \LbLcDsKK and \LbLcDs decays is $1.6\%$, which is assigned as the systematic uncertainty due to the hardware trigger efficiency correction.

The simulated PID response is calibrated using control samples and the kernel density estimation method~\cite{LHCb-PUB-2016-021,CRANMER2001198}. This correction accounts for two sources of uncertainty: one arising from the kernel density estimation algorithm and the other from the finite size of the calibration samples. To assess the systematic uncertainty, the kernel density estimation template is varied by increasing the kernel width by $50\%$ and by employing a bootstrapping method. The resulting variations in the efficiency ratios are assigned as systematic uncertainties corresponding to each source.

The systematic uncertainties related to the fit models are evaluated with pseudoexperiments generated with the baseline model and fitted with alternative signal and background parametrizations. For the \Lb signal, an alternative model is used, based on a modified Crystal Ball function with tails on both sides of the peak, while the background model is replaced with an exponential function instead of the Chebyshev polynomial. The yield shift is quantified as the difference between the yield obtained from the baseline fit to data and the mean yield derived from the ensemble of pseudoexperiments. Considering the shifts for both signal and normalization modes, the difference in the signal yield ratio is assigned as the systematic uncertainty.

To account for potential signal loss introduced by the random selection of multiple candidates, the procedure is repeated 500 times with different random seeds, performing a signal yield fit for each iteration. The corresponding systematic uncertainty is evaluated as the difference between the nominal signal yield and the mean of the resulting distribution.

The systematic uncertainty arising from the limited size of the simulated samples is estimated using a Bayesian approach, where the number of selected candidates for the true efficiency in each \mbox{$(m^2(\Lc\Kp\Km), m^2(\Dsm\Kp\Km))$} bin in Eq.~\eqref{equ:eff_corr_yield} is assumed to be binomially distributed, and the prior probability is chosen as a uniform distribution, both of which are used to determine the posterior probability of the efficiency in actual measurements. The uncertainty on the efficiency table is evaluated independently for each bin and subsequently propagated to the ratio of branching fractions. Additionally, the binning scheme used in the efficiency parametrization is varied, and the largest observed difference in the ratio of branching fractions is assigned as the systematic uncertainty.

To estimate the systematic uncertainty associated with the correction for \Lc and \Dsm decay amplitudes, the correction is removed from the simulated events. The resulting difference in the branching fraction ratio is assigned as a systematic uncertainty. The magnitude of this difference depends on the size of the simulated samples, making this uncertainty fully correlated with the uncertainty arising from the limited size of the simulated samples.

The summary of the systematic uncertainties is presented in Table~\ref{tab:Summary_Syst}, where correlations among different sources are taken into account when determining the total uncertainty. The systematic uncertainty on the relative branching fraction measurement is found to be smaller than the statistical uncertainty.

\begin{table}[tb]
    \caption{Summary of relative systematic uncertainties on the relative branching fraction.}
    \label{tab:Summary_Syst}
    \centering
    \begin{tabular}{cc}
        \hline
           Sources                                           & Relative uncertainty [$\%$]   \\
        \hline
            Fit bias                                         & $ 0.2 $ \\
            Tracking efficiency                              & $ 2.7 $ \\
            Trigger efficiency                               & $ 1.6 $ \\
            PID correction algorithm                         & $ 0.6 $ \\
            Signal fit model                                 & $ 0.6 $ \\
            Background fit model                             & $ 0.3 $ \\
            Multiple candidates                              & $ 2.5 $ \\
            Simulated sample sizes                           & $ 5.3 $ \\
            Binning scheme in $n_\textrm{corr}$ calculation  & $ 2.3 $ \\
           Correction on \Lc and \Dsm decay amplitudes       & $ 2.0 $ \\
        \hline
          Total                                              & $ 8.7 $ \\
        \hline 
    \end{tabular}
\end{table}

\section{Upper limits on the \texorpdfstring{$\Pcsz$}{P\_ccbars0} production rate}
\begin{figure}[tb]
    \centering
    \begin{minipage}[c]{0.45\textwidth}
        \centering
        \includegraphics[width=0.99\linewidth]{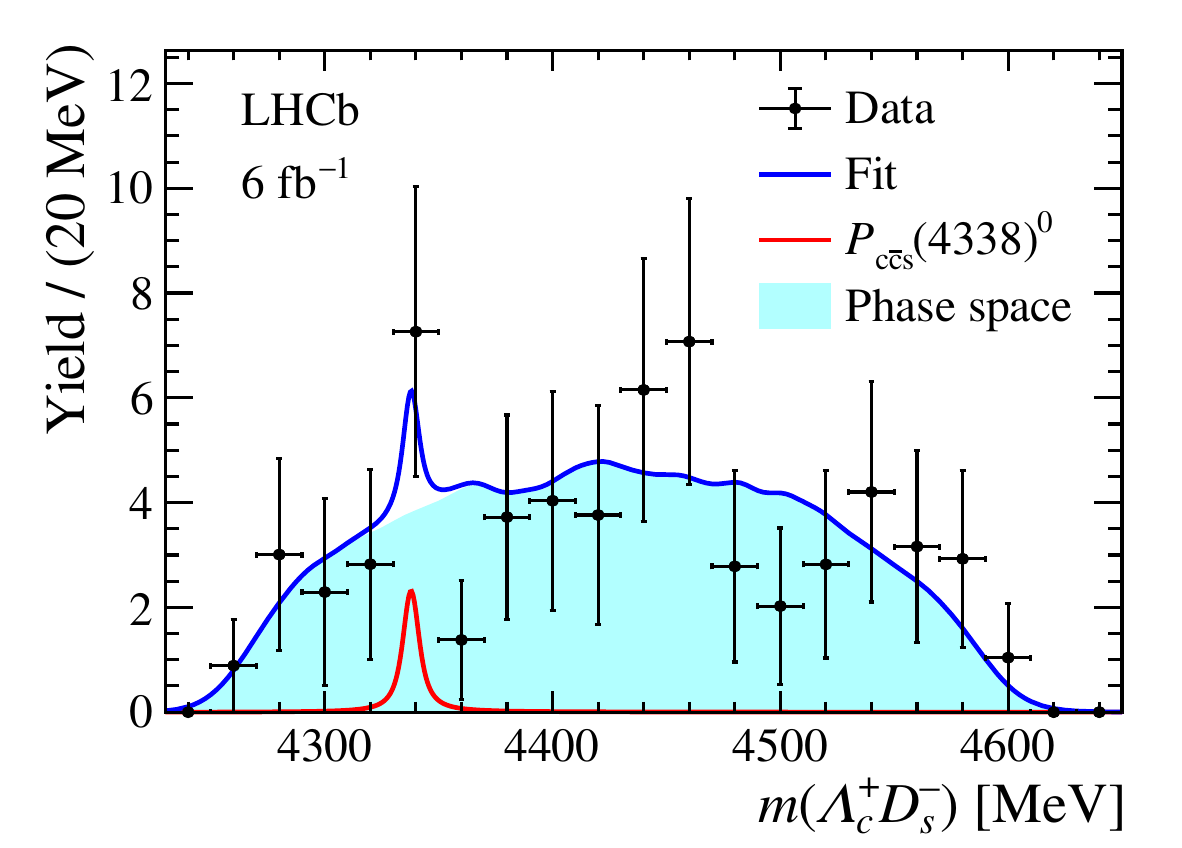}
    \end{minipage}
    \begin{minipage}[c]{0.45\textwidth}
        \centering
        \includegraphics[width=0.99\linewidth]{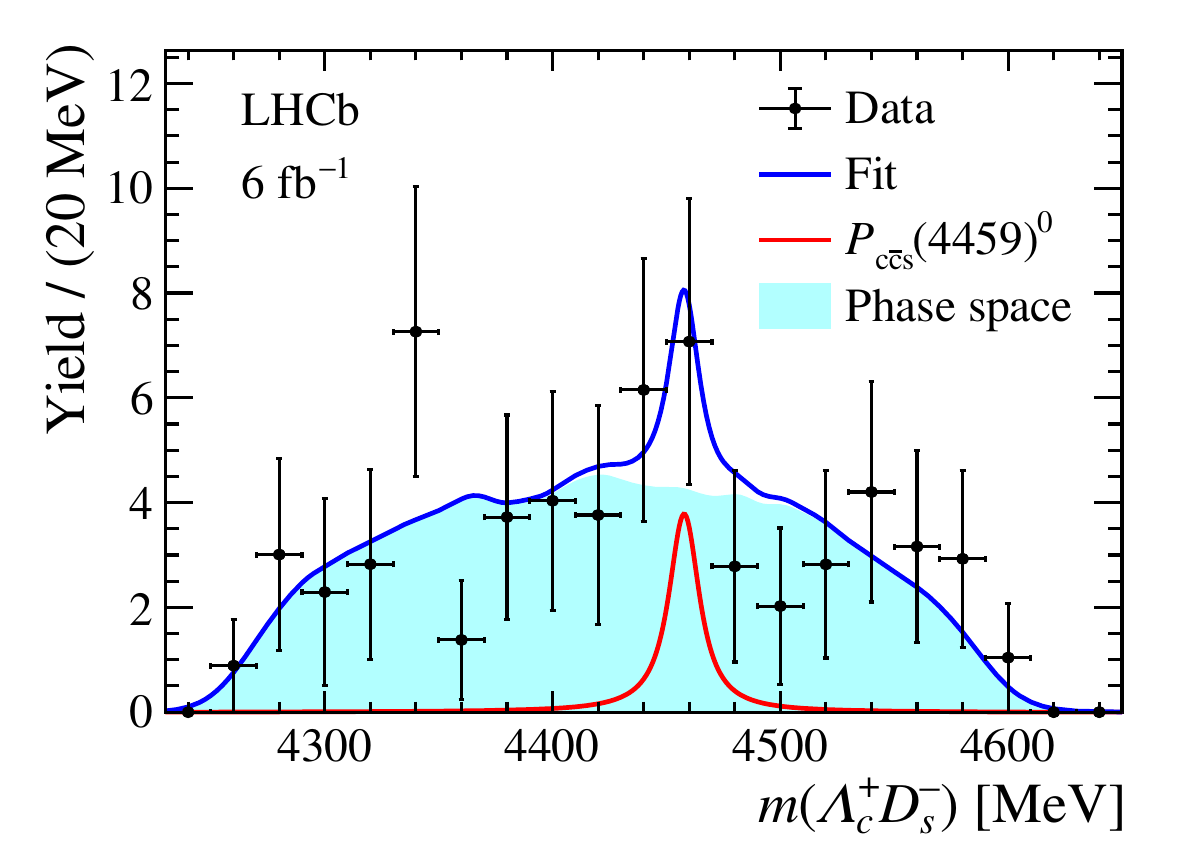}
    \end{minipage}
    \caption{Distribution of \mbox{$m(\Lc\Dsm)$} for background-subtracted \LbLcDsKK decays with the fit result also shown. The red line shows the contribution from the (left) $\Pcs(4338)^0$ and (right) $\Pcs(4459)^0$ state, while the cyan fill represents the phase-space distribution of the \LbLcDsPhi decay.}
    \label{fig:Pcs_fit_nominal}
\end{figure}

A search for a \decay{\Pcsz}{\Lc\Dsm} contribution to the \LbLcDsKK decay is performed by inspecting the mass distribution of \mbox{$\Lc\Dsm$} where the background is subtracted with the \sPlot technique. 
The mass \mbox{$m(\Lc\Dsm)$} is computed from a kinematic fit~\cite{Hulsbergen:2005pu} which requires the \Lb baryon to originate from its associated primary vertex and constrains the \Lc, \Dsm, and \Lb hadrons to their known masses~\cite{PDG2024}. 
A weighted unbinned maximum-likelihood fit~\cite{Xie:2009rka} is applied to the \mbox{$m(\Lc\Dsm)$} distribution, 
where the data are described as the incoherent sum of \decay{\Pcsz}{\Lc\Dsm} decays and a nonresonant \mbox{\Lc\Dsm} contribution.
The \Pcsz resonance is modeled using a relativistic Breit-Wigner function~\cite{Jackson:1964zd,PDG2024}, with parameters for $\Pcs(4338)^0$ and $\Pcs(4459)^0$ resonances obtained from Refs.~\cite{LHCb-PAPER-2022-031} and \cite{LHCb-PAPER-2020-039}, respectively. The line shape is described in detail in the Appendix. It is subsequently convolved with a Gaussian resolution function whose shape parameters are determined from simulation. The contribution from \LbLcDsKK decays with a nonresonant \mbox{\Lc\Dsm} system is modeled using simulated \LbLcDsPhi decays generated uniformly over the phase space. The fit projections for $\Pcs(4338)^0$ and $\Pcs(4459)^0$ resonances are shown in Fig.~\ref{fig:Pcs_fit_nominal}. Both fits yield significances below  $2\sigma$, showing no significant pentaquark contributions in the \mbox{\Lc\Dsm} system.

Defining the relative $\Pcsz$ contribution as
\begin{equation}
    \mathcal{R}_{\Pcsz} \equiv \RBF{\decay{\Lb}{\Pcs\Kp\Km}}{\LbLcDsKK} \cdot \BF\left(\decay{\Pcsz}{\Lc\Dsm}\right),
    \label{equ:ratio_Pcs}
\end{equation}
the upper limits on the production rates are determined to be \mbox{$\mathcal{R}_{\Pcs(4338)^0} < \MyRPcsLow  (0.10)$} and \mbox{$\mathcal{R}_{\Pcs(4459)^0} < \MyRPcsHigh  (0.17)$} at the $95\% \, (90\%)$ confidence level. These limits are derived by integrating the probability distribution of $ \mathcal{R}_{\Pcsz}$, ensuring that the integral up to the upper limit equals $0.95 \, (0.90)$ times the total integral~\cite{Narsky:1999kt}. Systematic uncertainties associated with the modeling of the pentaquark line shape and of the nonresonant contribution  are found to have negligible impact on the measured $\mathcal{R}_{\Pcsz}$ values.

\section{Conclusion}
The \LbLcDsKK decay has been observed for the first time, and  the branching fraction relative to the \LbLcDs decay is measured to be 
\begin{equation}
    \RBF{\LbLcDsKK}{\LbLcDs} = \RBRresultNoBrErr{\MyRBFNominal}{\MyRBFStat}{\MyRBFSyst}, \nonumber
    \label{equ:RBF_result}
\end{equation}
where the first uncertainty is statistical and the second is systematic.

No significant pentaquark signal was identified in the \mbox{$m(\Lc\Dsm)$} mass spectrum. With only 61 candidates detected, the sample size is insufficient to perform an amplitude analysis. Upper limits on the fit fractions of $\Pcs(4338)^0$ and $\Pcs(4459)^0$ decaying to the \mbox{$\Lc\Dsm$} final state are therefore set at $\MyRPcsLow  (0.10)$ and $\MyRPcsHigh  (0.17)$, respectively, at the $95\% \, (90\%)$ confidence level. Furthermore, no evidence for excited charmed hadrons or exotic states is found in the \mbox{$\Lc\phiz$} or \mbox{$\Dsm\phiz$} mass distributions.

It is anticipated that this statistical limitation will be surmounted with the ongoing higher-luminosity data acquisition, facilitated by the upgraded \lhcb detector. The larger data samples due to increased luminosity and trigger efficiency are expected to enable a more comprehensive analysis~\cite{LHCb-TDR-012,LHCb-DP-2022-002}. Future larger datasets will allow for a full amplitude analysis that can take into account the \Lc polarization~\cite{LHCb-PAPER-2022-002,LHCb-PAPER-2022-044} and $\phi$ decay degrees of freedom, which will contribute to the properties of the strange pentaquarks with hidden charm.

\clearpage

%% file: acknowledgements.tex
\section*{Acknowledgements}
%
%
\noindent We express our gratitude to our colleagues in the CERN
accelerator departments for the excellent performance of the LHC. We
thank the technical and administrative staff at the LHCb
institutes.
We acknowledge support from CERN and from the national agencies:
ARC (Australia);
CAPES, CNPq, FAPERJ and FINEP (Brazil); 
MOST and NSFC (China); 
CNRS/IN2P3 (France); 
BMBF, DFG and MPG (Germany); 
INFN (Italy); 
NWO (Netherlands); 
MNiSW and NCN (Poland); 
MCID/IFA (Romania); 
MICIU and AEI (Spain);
SNSF and SER (Switzerland); 
NASU (Ukraine); 
STFC (United Kingdom); 
DOE NP and NSF (USA).
We acknowledge the computing resources that are provided by ARDC (Australia), 
CBPF (Brazil),
CERN, 
IHEP and LZU (China),
IN2P3 (France), 
KIT and DESY (Germany), 
INFN (Italy), 
SURF (Netherlands),
Polish WLCG (Poland),
IFIN-HH (Romania), 
PIC (Spain), CSCS (Switzerland), 
and GridPP (United Kingdom).
We are indebted to the communities behind the multiple open-source
software packages on which we depend.
Individual groups or members have received support from
Key Research Program of Frontier Sciences of CAS, CAS PIFI, CAS CCEPP, 
Fundamental Research Funds for the Central Universities,  and Sci.\ \& Tech.\ Program of Guangzhou (China);
Minciencias (Colombia);
EPLANET, Marie Sk\l{}odowska-Curie Actions, ERC and NextGenerationEU (European Union);
A*MIDEX, ANR, IPhU and Labex P2IO, and R\'{e}gion Auvergne-Rh\^{o}ne-Alpes (France);
Alexander-von-Humboldt Foundation (Germany);
ICSC (Italy); 
Severo Ochoa and Mar\'ia de Maeztu Units of Excellence, GVA, XuntaGal, GENCAT, InTalent-Inditex and Prog.~Atracci\'on Talento CM (Spain);
SRC (Sweden);
the Leverhulme Trust, the Royal Society and UKRI (United Kingdom).

%% file: appendix.tex

\section*{Appendix}

\appendix
\numberwithin{equation}{section}

\section{Lineshape of the pentaquark} \label{app:lineshape}

For modeling the pentaquark resonances, the Breit-Wigner propagator is employed, expressed as
\begin{equation}
    \mathrm{BW}(m|m_0, \Gamma_0 ) = \frac{1}{m_0^{2} - m^2 - im_0\Gamma(m)}\,,
    \label{equ:RBW_propagator}
\end{equation}
where $m$ represents the mass of the $\mbox{\Lc \Dsm}$ system, $m_0$ represents the mass of the pentaquark taken from Ref.~\cite{PDG2024}, and $\Gamma(m)$ is the mass-dependent width. Given that the only known decay channel for $\Pcs(4338)^0$ and $\Pcs(4459)^0$ is $\mbox{\jpsi\Lz}$, the energy-dependent width is calculated based on the $\mbox{\jpsi \Lz}$ system. Under the $S$-wave assumption, the $\Gamma(m)$ is simplified as
\begin{equation}
    \Gamma(m) = \Gamma_0 \cdot \frac{q(m,m_\jpsi,m_\Lz)}{q(m_0,m_\jpsi,m_\Lz)} \cdot \frac{m_0}{m}\,,
    \label{equ:BW_running_width}
\end{equation}
where $\Gamma_0$ is the natural width of the pentaquark state, $q(m,m_1,m_2)$ is the breakup momentum calculated by
\begin{equation}
        q(m,m_1,m_2) = \frac{ \sqrt{\left[ m^2 - (m_1 + m_2)^2\right] \left[ m^2 - (m_1 - m_2)^2\right]} }{2m} \,,\label{equ:q_momentum}
\end{equation}
and $m_{1}$ and $m_{2}$ are the masses of the two decay products.

The pentaquark line shape $\mathcal{P}_\mathrm{sig}(m|m_0,\Gamma_0)$ is then obtained by incorporating a phase-space factor $\rho(m)=p_\Lb q_\Pcs$ and a normalization constant $\mathcal{N}$, derived as
\begin{equation}
    \mathcal{P}_\mathrm{sig}(m|m_0,\Gamma_0) = \frac{\left| \mathrm{BW}(m|m_0, \Gamma_0 )  \right|^2 \cdot  p_\Lb(m) \,q_\Pcs(m) }{\mathcal{N}},
\end{equation}
where $q_\Pcs(m)=q(m,m_\Lc,m_\Dsm)$ represents the momentum of \Lc in the pentaquark rest frame, and $p_\Lb(m)=q(m_\Lb,m,m_\phi)$ denotes the momentum of the pentaquark in the \Lb rest frame.

The normalized background line shape $\mathcal{P}_\mathrm{bkg}(m)$ is derived from the phase-space simulation~\cite{CRANMER2001198}, where the kinematics of the \Lb hadron and decay amplitudes of the \Lc and \Ds hadrons are corrected to match the data.

Finally, the total fit model is constructed by adding the signal and the background components, expressed as
\begin{equation}
    \mathcal{P}(m|m_0,\Gamma_0) = \mathcal{R}_\Pcsz \mathcal{P}_\mathrm{sig}(m|m_0,\Gamma_0)  +  (1-\mathcal{R}_\Pcsz)\mathcal{P}_\mathrm{bkg}(m).
    \label{equ:total_Pcs_model}
\end{equation}

%% file: Authorship_LHCb-PAPER-2025-022.tex
\centerline
{\large\bf LHCb collaboration}
\begin
{flushleft}
\small
R.~Aaij$^{38}$\lhcborcid{0000-0003-0533-1952},
A.S.W.~Abdelmotteleb$^{57}$\lhcborcid{0000-0001-7905-0542},
C.~Abellan~Beteta$^{51}$\lhcborcid{0009-0009-0869-6798},
F.~Abudin{\'e}n$^{57}$\lhcborcid{0000-0002-6737-3528},
T.~Ackernley$^{61}$\lhcborcid{0000-0002-5951-3498},
A. A. ~Adefisoye$^{69}$\lhcborcid{0000-0003-2448-1550},
B.~Adeva$^{47}$\lhcborcid{0000-0001-9756-3712},
M.~Adinolfi$^{55}$\lhcborcid{0000-0002-1326-1264},
P.~Adlarson$^{85}$\lhcborcid{0000-0001-6280-3851},
C.~Agapopoulou$^{14}$\lhcborcid{0000-0002-2368-0147}
% ...

{\footnotesize \it

% ...
$^{14}$Universit{\'e} Paris-Saclay, CNRS/IN2P3, IJCLab, Orsay, France\\
% ...
$^{38}$Nikhef National Institute for Subatomic Physics, Amsterdam, Netherlands\\
% ...
$^{47}$Instituto Galego de F{\'\i}sica de Altas Enerx{\'\i}as (IGFAE), Universidade de Santiago de Compostela, Santiago de Compostela, Spain\\
% ...
$^{51}$Physik-Institut, Universit{\"a}t Z{\"u}rich, Z{\"u}rich, Switzerland\\
% ...
$^{55}$H.H. Wills Physics Laboratory, University of Bristol, Bristol, United Kingdom\\
% ...
$^{57}$Department of Physics, University of Warwick, Coventry, United Kingdom\\
% ...
$^{60}$School of Physics and Astronomy, University of Glasgow, Glasgow, United Kingdom\\
$^{61}$Oliver Lodge Laboratory, University of Liverpool, Liverpool, United Kingdom\\
% ...
$^{69}$Syracuse University, Syracuse, NY, United States\\
% ...
$^{85}$Department of Physics and Astronomy, Uppsala University, Uppsala, Sweden, associated to $^{60}$\\
% ...
}
\end{flushleft}

%% file: main.bbl
\ifx\mcitethebibliography\mciteundefinedmacro
\PackageError{LHCb.bst}{mciteplus.sty has not been loaded}
{This bibstyle requires the use of the mciteplus package.}\fi
\providecommand{\href}[2]{#2}
\begin{mcitethebibliography}{10}
\mciteSetBstSublistMode{n}
\mciteSetBstMaxWidthForm{subitem}{\alph{mcitesubitemcount})}
\mciteSetBstSublistLabelBeginEnd{\mcitemaxwidthsubitemform\space}
{\relax}{\relax}

\bibitem{GellMann:1964nj}
M.~Gell-Mann, \ifthenelse{\boolean{articletitles}}{\emph{{A schematic model of baryons and mesons}}, }{}\href{https://doi.org/10.1016/S0031-9163(64)92001-3}{Phys.\ Lett.\  \textbf{8} (1964) 214}\relax
\mciteBstWouldAddEndPuncttrue
\mciteSetBstMidEndSepPunct{\mcitedefaultmidpunct}
{\mcitedefaultendpunct}{\mcitedefaultseppunct}\relax
\EndOfBibitem
\bibitem{Zweig:352337}
G.~Zweig, \ifthenelse{\boolean{articletitles}}{\emph{{An SU$_3$ model for strong interaction symmetry and its breaking; Version 1}}}{}  \href{http://cds.cern.ch/record/352337}{CERN-TH-401}, CERN, Geneva, 1964\relax
\mciteBstWouldAddEndPuncttrue
\mciteSetBstMidEndSepPunct{\mcitedefaultmidpunct}
{\mcitedefaultendpunct}{\mcitedefaultseppunct}\relax
\EndOfBibitem
\bibitem{LHCb-PAPER-2015-029}
LHCb collaboration, R.~Aaij {\em et~al.}, \ifthenelse{\boolean{articletitles}}{\emph{{Observation of $\jpsi\proton$ resonances consistent with pentaquark states in \mbox{\decay{\Lb}{\jpsi\proton\Km}} decays}}, }{}\href{https://doi.org/10.1103/PhysRevLett.115.072001}{Phys.\ Rev.\ Lett.\  \textbf{115} (2015) 072001}, \href{http://arxiv.org/abs/1507.03414}{{\normalfont\ttfamily arXiv:1507.03414}}\relax
\mciteBstWouldAddEndPuncttrue
\mciteSetBstMidEndSepPunct{\mcitedefaultmidpunct}
{\mcitedefaultendpunct}{\mcitedefaultseppunct}\relax
\EndOfBibitem
\bibitem{LHCb-PAPER-2016-009}
LHCb collaboration, R.~Aaij {\em et~al.}, \ifthenelse{\boolean{articletitles}}{\emph{{Model-independent evidence for $\jpsi\proton$ contributions to \mbox{\decay{\Lb}{\jpsi\proton\Km}} decays}}, }{}\href{https://doi.org/10.1103/PhysRevLett.117.082002}{Phys.\ Rev.\ Lett.\  \textbf{117} (2016) 082002}, \href{http://arxiv.org/abs/1604.05708}{{\normalfont\ttfamily arXiv:1604.05708}}\relax
\mciteBstWouldAddEndPuncttrue
\mciteSetBstMidEndSepPunct{\mcitedefaultmidpunct}
{\mcitedefaultendpunct}{\mcitedefaultseppunct}\relax
\EndOfBibitem
\bibitem{LHCb-PAPER-2019-014}
LHCb collaboration, R.~Aaij {\em et~al.}, \ifthenelse{\boolean{articletitles}}{\emph{{Observation of a narrow pentaquark state, $P_c(4312)^+$, and of two-peak structure of the $P_c(4450)^+$}}, }{}\href{https://doi.org/10.1103/PhysRevLett.122.222001}{Phys.\ Rev.\ Lett.\  \textbf{122} (2019) 222001}, \href{http://arxiv.org/abs/1904.03947}{{\normalfont\ttfamily arXiv:1904.03947}}\relax
\mciteBstWouldAddEndPuncttrue
\mciteSetBstMidEndSepPunct{\mcitedefaultmidpunct}
{\mcitedefaultendpunct}{\mcitedefaultseppunct}\relax
\EndOfBibitem
\bibitem{LHCb-PAPER-2022-031}
LHCb collaboration, R.~Aaij {\em et~al.}, \ifthenelse{\boolean{articletitles}}{\emph{{Observation of a $\jpsi\Lambdares$ resonance consistent with a strange pentaquark candidate in $\Bm\to \jpsi\Lambdares\antiproton$ decays}}, }{}\href{https://doi.org/10.1103/PhysRevLett.131.031901}{Phys.\ Rev.\ Lett.\  \textbf{131} (2023) 031901}, \href{http://arxiv.org/abs/2210.10346}{{\normalfont\ttfamily arXiv:2210.10346}}\relax
\mciteBstWouldAddEndPuncttrue
\mciteSetBstMidEndSepPunct{\mcitedefaultmidpunct}
{\mcitedefaultendpunct}{\mcitedefaultseppunct}\relax
\EndOfBibitem
\bibitem{LHCb-PAPER-2021-018}
LHCb collaboration, R.~Aaij {\em et~al.}, \ifthenelse{\boolean{articletitles}}{\emph{{Evidence for a new structure in the $\jpsi \proton$ and $\jpsi \antiproton$ systems in $\Bs \to \jpsi\proton\antiproton$ decays}}, }{}\href{https://doi.org/10.1103/PhysRevLett.128.062001}{Phys.\ Rev.\ Lett.\  \textbf{128} (2022) 062001}, \href{http://arxiv.org/abs/2108.04720}{{\normalfont\ttfamily arXiv:2108.04720}}\relax
\mciteBstWouldAddEndPuncttrue
\mciteSetBstMidEndSepPunct{\mcitedefaultmidpunct}
{\mcitedefaultendpunct}{\mcitedefaultseppunct}\relax
\EndOfBibitem
\bibitem{LHCb-PAPER-2020-039}
LHCb collaboration, R.~Aaij {\em et~al.}, \ifthenelse{\boolean{articletitles}}{\emph{{Evidence of a $\jpsi\Lz$ structure and observation of excited $\Xim$ states in the $\Xibm \rightarrow \jpsi \Lz \Km$ decay}}, }{}\href{https://doi.org/10.1016/j.scib.2021.02.030}{Science Bulletin \textbf{66} (2021) 1278}, \href{http://arxiv.org/abs/2012.10380}{{\normalfont\ttfamily arXiv:2012.10380}}\relax
\mciteBstWouldAddEndPuncttrue
\mciteSetBstMidEndSepPunct{\mcitedefaultmidpunct}
{\mcitedefaultendpunct}{\mcitedefaultseppunct}\relax
\EndOfBibitem
\bibitem{Wu:2010jy}
J.-J. Wu, R.~Molina, E.~Oset, and B.~S. Zou, \ifthenelse{\boolean{articletitles}}{\emph{{Prediction of narrow $N^*$ and $\Lambda^*$ resonances with hidden charm above 4 GeV}}, }{}\href{https://doi.org/10.1103/PhysRevLett.105.232001}{Phys.\ Rev.\ Lett.\  \textbf{105} (2010) 232001}, \href{http://arxiv.org/abs/1007.0573}{{\normalfont\ttfamily arXiv:1007.0573}}\relax
\mciteBstWouldAddEndPuncttrue
\mciteSetBstMidEndSepPunct{\mcitedefaultmidpunct}
{\mcitedefaultendpunct}{\mcitedefaultseppunct}\relax
\EndOfBibitem
\bibitem{Wu:2010vk}
J.-J. Wu, R.~Molina, E.~Oset, and B.~S. Zou, \ifthenelse{\boolean{articletitles}}{\emph{{Dynamically generated $N^{*}$ and $\Lambda^*$ resonances in the hidden charm sector around 4.3 GeV}}, }{}\href{https://doi.org/10.1103/PhysRevC.84.015202}{Phys.\ Rev.\  \textbf{C84} (2011) 015202}, \href{http://arxiv.org/abs/1011.2399}{{\normalfont\ttfamily arXiv:1011.2399}}\relax
\mciteBstWouldAddEndPuncttrue
\mciteSetBstMidEndSepPunct{\mcitedefaultmidpunct}
{\mcitedefaultendpunct}{\mcitedefaultseppunct}\relax
\EndOfBibitem
\bibitem{Yang:2011wz}
Z.-C. Yang {\em et~al.}, \ifthenelse{\boolean{articletitles}}{\emph{{The possible hidden-charm molecular baryons composed of anti-charmed meson and charmed baryon}}, }{}\href{https://doi.org/10.1088/1674-1137/36/1/002}{Chin.\ Phys.\  \textbf{C36} (2012) 6}, \href{http://arxiv.org/abs/1105.2901}{{\normalfont\ttfamily arXiv:1105.2901}}\relax
\mciteBstWouldAddEndPuncttrue
\mciteSetBstMidEndSepPunct{\mcitedefaultmidpunct}
{\mcitedefaultendpunct}{\mcitedefaultseppunct}\relax
\EndOfBibitem
\bibitem{Dong:2021juy}
X.-K. Dong, F.-K. Guo, and B.-S. Zou, \ifthenelse{\boolean{articletitles}}{\emph{{A survey of heavy-antiheavy hadronic molecules}}, }{}\href{https://doi.org/10.13725/j.cnki.pip.2021.02.001}{Progr.\ Phys.\  \textbf{41} (2021) 65}, \href{http://arxiv.org/abs/2101.01021}{{\normalfont\ttfamily arXiv:2101.01021}}\relax
\mciteBstWouldAddEndPuncttrue
\mciteSetBstMidEndSepPunct{\mcitedefaultmidpunct}
{\mcitedefaultendpunct}{\mcitedefaultseppunct}\relax
\EndOfBibitem
\bibitem{Guo:2017jvc}
F.-K. Guo {\em et~al.}, \ifthenelse{\boolean{articletitles}}{\emph{{Hadronic molecules}}, }{}\href{https://doi.org/10.1103/RevModPhys.90.015004}{Rev.\ Mod.\ Phys.\  \textbf{90} (2018) 015004}, Erratum \href{https://doi.org/}{ibid.\   \textbf{94} (2022) 029901}, \href{http://arxiv.org/abs/1705.00141}{{\normalfont\ttfamily arXiv:1705.00141}}\relax
\mciteBstWouldAddEndPuncttrue
\mciteSetBstMidEndSepPunct{\mcitedefaultmidpunct}
{\mcitedefaultendpunct}{\mcitedefaultseppunct}\relax
\EndOfBibitem
\bibitem{Yuan:2012wz}
S.~G. Yuan {\em et~al.}, \ifthenelse{\boolean{articletitles}}{\emph{{Study of $qqqc\bar{c}$ five quark system with three kinds of quark-quark hyperfine interaction}}, }{}\href{https://doi.org/10.1140/epja/i2012-12061-2}{Eur.\ Phys.\ J.\  \textbf{A48} (2012) 61}, \href{http://arxiv.org/abs/1201.0807}{{\normalfont\ttfamily arXiv:1201.0807}}\relax
\mciteBstWouldAddEndPuncttrue
\mciteSetBstMidEndSepPunct{\mcitedefaultmidpunct}
{\mcitedefaultendpunct}{\mcitedefaultseppunct}\relax
\EndOfBibitem
\bibitem{Maiani:2015vwa}
L.~Maiani, A.~D. Polosa, and V.~Riquer, \ifthenelse{\boolean{articletitles}}{\emph{{The new pentaquarks in the diquark model}}, }{}\href{https://doi.org/10.1016/j.physletb.2015.08.008}{Phys.\ Lett.\  \textbf{B749} (2015) 289}, \href{http://arxiv.org/abs/1507.04980}{{\normalfont\ttfamily arXiv:1507.04980}}\relax
\mciteBstWouldAddEndPuncttrue
\mciteSetBstMidEndSepPunct{\mcitedefaultmidpunct}
{\mcitedefaultendpunct}{\mcitedefaultseppunct}\relax
\EndOfBibitem
\bibitem{Lebed:2015tna}
R.~F. Lebed, \ifthenelse{\boolean{articletitles}}{\emph{{The pentaquark candidates in the dynamical diquark picture}}, }{}\href{https://doi.org/10.1016/j.physletb.2015.08.032}{Phys.\ Lett.\  \textbf{D749} (2015) 454}, \href{http://arxiv.org/abs/1507.05867}{{\normalfont\ttfamily arXiv:1507.05867}}\relax
\mciteBstWouldAddEndPuncttrue
\mciteSetBstMidEndSepPunct{\mcitedefaultmidpunct}
{\mcitedefaultendpunct}{\mcitedefaultseppunct}\relax
\EndOfBibitem
\bibitem{Li:2015gta}
G.-N. Li, X.-G. He, and M.~He, \ifthenelse{\boolean{articletitles}}{\emph{{Some predictions of diquark model for hidden charm pentaquark discovered at the LHCb}}, }{}\href{https://doi.org/10.1007/JHEP12(2015)128}{JHEP \textbf{12} (2015) 128}, \href{http://arxiv.org/abs/1507.08252}{{\normalfont\ttfamily arXiv:1507.08252}}\relax
\mciteBstWouldAddEndPuncttrue
\mciteSetBstMidEndSepPunct{\mcitedefaultmidpunct}
{\mcitedefaultendpunct}{\mcitedefaultseppunct}\relax
\EndOfBibitem
\bibitem{Esposito:2016noz}
A.~Esposito, A.~Pilloni, and A.~D. Polosa, \ifthenelse{\boolean{articletitles}}{\emph{{Multiquark resonances}}, }{}\href{https://doi.org/10.1016/j.physrep.2016.11.002}{Phys.\ Rept.\  \textbf{668} (2017) 1}, \href{http://arxiv.org/abs/1611.07920}{{\normalfont\ttfamily arXiv:1611.07920}}\relax
\mciteBstWouldAddEndPuncttrue
\mciteSetBstMidEndSepPunct{\mcitedefaultmidpunct}
{\mcitedefaultendpunct}{\mcitedefaultseppunct}\relax
\EndOfBibitem
\bibitem{Richard:2016eis}
J.-M. Richard, \ifthenelse{\boolean{articletitles}}{\emph{{Exotic hadrons: review and perspectives}}, }{}\href{https://doi.org/10.1007/s00601-016-1159-0}{Few Body Syst.\  \textbf{57} (2016) 1185}, \href{http://arxiv.org/abs/1606.08593}{{\normalfont\ttfamily arXiv:1606.08593}}\relax
\mciteBstWouldAddEndPuncttrue
\mciteSetBstMidEndSepPunct{\mcitedefaultmidpunct}
{\mcitedefaultendpunct}{\mcitedefaultseppunct}\relax
\EndOfBibitem
\bibitem{Zhu:2015bba}
R.~Zhu and C.-F. Qiao, \ifthenelse{\boolean{articletitles}}{\emph{{Pentaquark states in a diquark\textendash{}triquark model}}, }{}\href{https://doi.org/10.1016/j.physletb.2016.03.022}{Phys.\ Lett.\  \textbf{B756} (2016) 259}, \href{http://arxiv.org/abs/1510.08693}{{\normalfont\ttfamily arXiv:1510.08693}}\relax
\mciteBstWouldAddEndPuncttrue
\mciteSetBstMidEndSepPunct{\mcitedefaultmidpunct}
{\mcitedefaultendpunct}{\mcitedefaultseppunct}\relax
\EndOfBibitem
\bibitem{Guo:2015umn}
F.-K. Guo, U.-G. Mei\ss{}ner, W.~Wang, and Z.~Yang, \ifthenelse{\boolean{articletitles}}{\emph{{How to reveal the exotic nature of the P$_c$(4450)}}, }{}\href{https://doi.org/10.1103/PhysRevD.92.071502}{Phys.\ Rev.\  \textbf{D92} (2015) 071502}, \href{http://arxiv.org/abs/1507.04950}{{\normalfont\ttfamily arXiv:1507.04950}}\relax
\mciteBstWouldAddEndPuncttrue
\mciteSetBstMidEndSepPunct{\mcitedefaultmidpunct}
{\mcitedefaultendpunct}{\mcitedefaultseppunct}\relax
\EndOfBibitem
\bibitem{Liu:2015fea}
X.-H. Liu, Q.~Wang, and Q.~Zhao, \ifthenelse{\boolean{articletitles}}{\emph{{Understanding the newly observed heavy pentaquark candidates}}, }{}\href{https://doi.org/10.1016/j.physletb.2016.03.089}{Phys.\ Lett.\  \textbf{B757} (2016) 231}, \href{http://arxiv.org/abs/1507.05359}{{\normalfont\ttfamily arXiv:1507.05359}}\relax
\mciteBstWouldAddEndPuncttrue
\mciteSetBstMidEndSepPunct{\mcitedefaultmidpunct}
{\mcitedefaultendpunct}{\mcitedefaultseppunct}\relax
\EndOfBibitem
\bibitem{Bayar:2016ftu}
M.~Bayar, F.~Aceti, F.-K. Guo, and E.~Oset, \ifthenelse{\boolean{articletitles}}{\emph{{A discussion on triangle singularities in the $\Lambda_b \to J/\psi K^{-} p$ reaction}}, }{}\href{https://doi.org/10.1103/PhysRevD.94.074039}{Phys.\ Rev.\  \textbf{D94} (2016) 074039}, \href{http://arxiv.org/abs/1609.04133}{{\normalfont\ttfamily arXiv:1609.04133}}\relax
\mciteBstWouldAddEndPuncttrue
\mciteSetBstMidEndSepPunct{\mcitedefaultmidpunct}
{\mcitedefaultendpunct}{\mcitedefaultseppunct}\relax
\EndOfBibitem
\bibitem{Guo:2019twa}
F.-K. Guo, X.-H. Liu, and S.~Sakai, \ifthenelse{\boolean{articletitles}}{\emph{{Threshold cusps and triangle singularities in hadronic reactions}}, }{}\href{https://doi.org/10.1016/j.ppnp.2020.103757}{Prog.\ Part.\ Nucl.\ Phys.\  \textbf{112} (2020) 103757}, \href{http://arxiv.org/abs/1912.07030}{{\normalfont\ttfamily arXiv:1912.07030}}\relax
\mciteBstWouldAddEndPuncttrue
\mciteSetBstMidEndSepPunct{\mcitedefaultmidpunct}
{\mcitedefaultendpunct}{\mcitedefaultseppunct}\relax
\EndOfBibitem
\bibitem{Chen:2016qju}
H.-X. Chen, W.~Chen, X.~Liu, and S.-L. Zhu, \ifthenelse{\boolean{articletitles}}{\emph{{The hidden-charm pentaquark and tetraquark states}}, }{}\href{https://doi.org/10.1016/j.physrep.2016.05.004}{Phys.\ Rept.\  \textbf{639} (2016) 1}, \href{http://arxiv.org/abs/1601.02092}{{\normalfont\ttfamily arXiv:1601.02092}}\relax
\mciteBstWouldAddEndPuncttrue
\mciteSetBstMidEndSepPunct{\mcitedefaultmidpunct}
{\mcitedefaultendpunct}{\mcitedefaultseppunct}\relax
\EndOfBibitem
\bibitem{Dong:2017gaw}
Y.~Dong, A.~Faessler, and V.~E. Lyubovitskij, \ifthenelse{\boolean{articletitles}}{\emph{{Description of heavy exotic resonances as molecular states using phenomenological Lagrangians}}, }{}\href{https://doi.org/10.1016/j.ppnp.2017.01.002}{Prog.\ Part.\ Nucl.\ Phys.\  \textbf{94} (2017) 282}\relax
\mciteBstWouldAddEndPuncttrue
\mciteSetBstMidEndSepPunct{\mcitedefaultmidpunct}
{\mcitedefaultendpunct}{\mcitedefaultseppunct}\relax
\EndOfBibitem
\bibitem{Olsen:2017bmm}
S.~L. Olsen, T.~Skwarnicki, and D.~Zieminska, \ifthenelse{\boolean{articletitles}}{\emph{{Nonstandard heavy mesons and baryons: Experimental evidence}}, }{}\href{https://doi.org/10.1103/RevModPhys.90.015003}{Rev.\ Mod.\ Phys.\  \textbf{90} (2018) 015003}, \href{http://arxiv.org/abs/1708.04012}{{\normalfont\ttfamily arXiv:1708.04012}}\relax
\mciteBstWouldAddEndPuncttrue
\mciteSetBstMidEndSepPunct{\mcitedefaultmidpunct}
{\mcitedefaultendpunct}{\mcitedefaultseppunct}\relax
\EndOfBibitem
\bibitem{LHCB-PAPER-2023-018}
LHCb collaboration, R.~Aaij {\em et~al.}, \ifthenelse{\boolean{articletitles}}{\emph{{Search for prompt production of pentaquarks in open charm hadron final states}}, }{}\href{https://doi.org/10.1103/PhysRevD.110.032001}{{Phys.\ Rev.\ } \textbf{D110} (2024) 032001}, \href{http://arxiv.org/abs/2404.07131}{{\normalfont\ttfamily arXiv:2404.07131}}\relax
\mciteBstWouldAddEndPuncttrue
\mciteSetBstMidEndSepPunct{\mcitedefaultmidpunct}
{\mcitedefaultendpunct}{\mcitedefaultseppunct}\relax
\EndOfBibitem
\bibitem{LHCb-PAPER-2023-034}
LHCb collaboration, R.~Aaij {\em et~al.}, \ifthenelse{\boolean{articletitles}}{\emph{{Observation of $\Lb \to \Lc \bar{D}^{(*)0} \Km $ and \mbox{$\Lb \to \Lc \Dssm$} decays}}, }{}\href{https://doi.org/10.1140/epjc/s10052-024-12752-3}{{Eur.\ Phys.\ J.\ } \textbf{C84} (2024) 575}, \href{http://arxiv.org/abs/2311.14088}{{\normalfont\ttfamily arXiv:2311.14088}}\relax
\mciteBstWouldAddEndPuncttrue
\mciteSetBstMidEndSepPunct{\mcitedefaultmidpunct}
{\mcitedefaultendpunct}{\mcitedefaultseppunct}\relax
\EndOfBibitem
\bibitem{LHCb-paper-2023-044}
LHCb collaboration, R.~Aaij {\em et~al.}, \ifthenelse{\boolean{articletitles}}{\emph{{First observation of $\Lambda_b^0 \to \Sigma_c^{(*)++} D^{(*)-} K^-$ decays}}, }{}\href{https://doi.org/10.1103/PhysRevD.110.L031104}{{Phys.\ Rev.\ } \textbf{D110} (2024) L031104}, \href{http://arxiv.org/abs/2404.19510}{{\normalfont\ttfamily arXiv:2404.19510}}\relax
\mciteBstWouldAddEndPuncttrue
\mciteSetBstMidEndSepPunct{\mcitedefaultmidpunct}
{\mcitedefaultendpunct}{\mcitedefaultseppunct}\relax
\EndOfBibitem
\bibitem{LHCb-DP-2008-001}
LHCb collaboration, A.~A. Alves~Jr.\ {\em et~al.}, \ifthenelse{\boolean{articletitles}}{\emph{{The \lhcb detector at the LHC}}, }{}\href{https://doi.org/10.1088/1748-0221/3/08/S08005}{JINST \textbf{3} (2008) S08005}\relax
\mciteBstWouldAddEndPuncttrue
\mciteSetBstMidEndSepPunct{\mcitedefaultmidpunct}
{\mcitedefaultendpunct}{\mcitedefaultseppunct}\relax
\EndOfBibitem
\bibitem{LHCb-DP-2014-002}
LHCb collaboration, R.~Aaij {\em et~al.}, \ifthenelse{\boolean{articletitles}}{\emph{{LHCb detector performance}}, }{}\href{https://doi.org/10.1142/S0217751X15300227}{Int.\ J.\ Mod.\ Phys.\  \textbf{A30} (2015) 1530022}, \href{http://arxiv.org/abs/1412.6352}{{\normalfont\ttfamily arXiv:1412.6352}}\relax
\mciteBstWouldAddEndPuncttrue
\mciteSetBstMidEndSepPunct{\mcitedefaultmidpunct}
{\mcitedefaultendpunct}{\mcitedefaultseppunct}\relax
\EndOfBibitem
\bibitem{LHCb-DP-2012-004}
R.~Aaij {\em et~al.}, \ifthenelse{\boolean{articletitles}}{\emph{{The \lhcb trigger and its performance in 2011}}, }{}\href{https://doi.org/10.1088/1748-0221/8/04/P04022}{JINST \textbf{8} (2013) P04022}, \href{http://arxiv.org/abs/1211.3055}{{\normalfont\ttfamily arXiv:1211.3055}}\relax
\mciteBstWouldAddEndPuncttrue
\mciteSetBstMidEndSepPunct{\mcitedefaultmidpunct}
{\mcitedefaultendpunct}{\mcitedefaultseppunct}\relax
\EndOfBibitem
\bibitem{BBDT}
V.~V. Gligorov and M.~Williams, \ifthenelse{\boolean{articletitles}}{\emph{{Efficient, reliable and fast high-level triggering using a bonsai boosted decision tree}}, }{}\href{https://doi.org/10.1088/1748-0221/8/02/P02013}{JINST \textbf{8} (2013) P02013}, \href{http://arxiv.org/abs/1210.6861}{{\normalfont\ttfamily arXiv:1210.6861}}\relax
\mciteBstWouldAddEndPuncttrue
\mciteSetBstMidEndSepPunct{\mcitedefaultmidpunct}
{\mcitedefaultendpunct}{\mcitedefaultseppunct}\relax
\EndOfBibitem
\bibitem{LHCb-PROC-2015-018}
T.~Likhomanenko {\em et~al.}, \ifthenelse{\boolean{articletitles}}{\emph{{LHCb topological trigger reoptimization}}, }{}\href{https://doi.org/10.1088/1742-6596/664/8/082025}{J.\ Phys.\ Conf.\ Ser.\  \textbf{664} (2015) 082025}, \href{http://arxiv.org/abs/1510.00572}{{\normalfont\ttfamily arXiv:1510.00572}}\relax
\mciteBstWouldAddEndPuncttrue
\mciteSetBstMidEndSepPunct{\mcitedefaultmidpunct}
{\mcitedefaultendpunct}{\mcitedefaultseppunct}\relax
\EndOfBibitem
\bibitem{Sjostrand:2007gs}
T.~Sj\"{o}strand, S.~Mrenna, and P.~Skands, \ifthenelse{\boolean{articletitles}}{\emph{{A brief introduction to PYTHIA 8.1}}, }{}\href{https://doi.org/10.1016/j.cpc.2008.01.036}{Comput.\ Phys.\ Commun.\  \textbf{178} (2008) 852}, \href{http://arxiv.org/abs/0710.3820}{{\normalfont\ttfamily arXiv:0710.3820}}\relax
\mciteBstWouldAddEndPuncttrue
\mciteSetBstMidEndSepPunct{\mcitedefaultmidpunct}
{\mcitedefaultendpunct}{\mcitedefaultseppunct}\relax
\EndOfBibitem
\bibitem{Sjostrand:2006za}
T.~Sj\"{o}strand, S.~Mrenna, and P.~Skands, \ifthenelse{\boolean{articletitles}}{\emph{{PYTHIA 6.4 physics and manual}}, }{}\href{https://doi.org/10.1088/1126-6708/2006/05/026}{JHEP \textbf{05} (2006) 026}, \href{http://arxiv.org/abs/hep-ph/0603175}{{\normalfont\ttfamily arXiv:hep-ph/0603175}}\relax
\mciteBstWouldAddEndPuncttrue
\mciteSetBstMidEndSepPunct{\mcitedefaultmidpunct}
{\mcitedefaultendpunct}{\mcitedefaultseppunct}\relax
\EndOfBibitem
\bibitem{LHCb-PROC-2010-056}
I.~Belyaev {\em et~al.}, \ifthenelse{\boolean{articletitles}}{\emph{{Handling of the generation of primary events in Gauss, the LHCb simulation framework}}, }{}\href{https://doi.org/10.1088/1742-6596/331/3/032047}{J.\ Phys.\ Conf.\ Ser.\  \textbf{331} (2011) 032047}\relax
\mciteBstWouldAddEndPuncttrue
\mciteSetBstMidEndSepPunct{\mcitedefaultmidpunct}
{\mcitedefaultendpunct}{\mcitedefaultseppunct}\relax
\EndOfBibitem
\bibitem{Lange:2001uf}
D.~J. Lange, \ifthenelse{\boolean{articletitles}}{\emph{{The EvtGen particle decay simulation package}}, }{}\href{https://doi.org/10.1016/S0168-9002(01)00089-4}{Nucl.\ Instrum.\ Meth.\  \textbf{A462} (2001) 152}\relax
\mciteBstWouldAddEndPuncttrue
\mciteSetBstMidEndSepPunct{\mcitedefaultmidpunct}
{\mcitedefaultendpunct}{\mcitedefaultseppunct}\relax
\EndOfBibitem
\bibitem{davidson2015photos}
N.~Davidson, T.~Przedzinski, and Z.~Was, \ifthenelse{\boolean{articletitles}}{\emph{{PHOTOS interface in C++: Technical and physics documentation}}, }{}\href{https://doi.org/https://doi.org/10.1016/j.cpc.2015.09.013}{Comp.\ Phys.\ Comm.\  \textbf{199} (2016) 86}, \href{http://arxiv.org/abs/1011.0937}{{\normalfont\ttfamily arXiv:1011.0937}}\relax
\mciteBstWouldAddEndPuncttrue
\mciteSetBstMidEndSepPunct{\mcitedefaultmidpunct}
{\mcitedefaultendpunct}{\mcitedefaultseppunct}\relax
\EndOfBibitem
\bibitem{Allison:2006ve}
Geant4 collaboration, J.~Allison {\em et~al.}, \ifthenelse{\boolean{articletitles}}{\emph{{Geant4 developments and applications}}, }{}\href{https://doi.org/10.1109/TNS.2006.869826}{IEEE Trans.\ Nucl.\ Sci.\  \textbf{53} (2006) 270}\relax
\mciteBstWouldAddEndPuncttrue
\mciteSetBstMidEndSepPunct{\mcitedefaultmidpunct}
{\mcitedefaultendpunct}{\mcitedefaultseppunct}\relax
\EndOfBibitem
\bibitem{Agostinelli:2002hh}
Geant4 collaboration, S.~Agostinelli {\em et~al.}, \ifthenelse{\boolean{articletitles}}{\emph{{Geant4: A simulation toolkit}}, }{}\href{https://doi.org/10.1016/S0168-9002(03)01368-8}{Nucl.\ Instrum.\ Meth.\  \textbf{A506} (2003) 250}\relax
\mciteBstWouldAddEndPuncttrue
\mciteSetBstMidEndSepPunct{\mcitedefaultmidpunct}
{\mcitedefaultendpunct}{\mcitedefaultseppunct}\relax
\EndOfBibitem
\bibitem{LHCb-PROC-2011-006}
M.~Clemencic {\em et~al.}, \ifthenelse{\boolean{articletitles}}{\emph{{The \lhcb simulation application, Gauss: Design, evolution and experience}}, }{}\href{https://doi.org/10.1088/1742-6596/331/3/032023}{J.\ Phys.\ Conf.\ Ser.\  \textbf{331} (2011) 032023}\relax
\mciteBstWouldAddEndPuncttrue
\mciteSetBstMidEndSepPunct{\mcitedefaultmidpunct}
{\mcitedefaultendpunct}{\mcitedefaultseppunct}\relax
\EndOfBibitem
\bibitem{LHCb-DP-2023-004}
LHCb collaboration, R.~Aaij {\em et~al.}, \ifthenelse{\boolean{articletitles}}{\emph{{Tracking of charged particles with nanosecond lifetimes at LHCb}}, }{}\href{https://doi.org/10.1140/epjc/s10052-024-12993-2}{Eur.\ Phys.\ J.\  \textbf{C84} (2024) 761}, \href{http://arxiv.org/abs/2403.09483}{{\normalfont\ttfamily arXiv:2403.09483}}\relax
\mciteBstWouldAddEndPuncttrue
\mciteSetBstMidEndSepPunct{\mcitedefaultmidpunct}
{\mcitedefaultendpunct}{\mcitedefaultseppunct}\relax
\EndOfBibitem
\bibitem{GBDT}
J.~H. Friedman, \ifthenelse{\boolean{articletitles}}{\emph{{Greedy function approximation: A gradient boosting machine}}, }{}\href{https://doi.org/10.1214/aos/1013203451}{Annals Statist.\  \textbf{29} (2001) 1189 }\relax
\mciteBstWouldAddEndPuncttrue
\mciteSetBstMidEndSepPunct{\mcitedefaultmidpunct}
{\mcitedefaultendpunct}{\mcitedefaultseppunct}\relax
\EndOfBibitem
\bibitem{Hocker:2007ht}
H.~Voss, A.~Hoecker, J.~Stelzer, and F.~Tegenfeldt, \ifthenelse{\boolean{articletitles}}{\emph{{TMVA - Toolkit for Multivariate Data Analysis with ROOT}}, }{}\href{https://doi.org/10.22323/1.050.0040}{PoS \textbf{ACAT} (2007) 040}\relax
\mciteBstWouldAddEndPuncttrue
\mciteSetBstMidEndSepPunct{\mcitedefaultmidpunct}
{\mcitedefaultendpunct}{\mcitedefaultseppunct}\relax
\EndOfBibitem
\bibitem{LHCb-PAPER-2014-002}
LHCb collaboration, R.~Aaij {\em et~al.}, \ifthenelse{\boolean{articletitles}}{\emph{{Study of beauty hadron decays into pairs of charm hadrons}}, }{}\href{https://doi.org/10.1103/PhysRevLett.112.202001}{Phys.\ Rev.\ Lett.\  \textbf{112} (2014) 202001}, \href{http://arxiv.org/abs/1403.3606}{{\normalfont\ttfamily arXiv:1403.3606}}\relax
\mciteBstWouldAddEndPuncttrue
\mciteSetBstMidEndSepPunct{\mcitedefaultmidpunct}
{\mcitedefaultendpunct}{\mcitedefaultseppunct}\relax
\EndOfBibitem
\bibitem{Hulsbergen:2005pu}
W.~D. Hulsbergen, \ifthenelse{\boolean{articletitles}}{\emph{{Decay chain fitting with a Kalman filter}}, }{}\href{https://doi.org/10.1016/j.nima.2005.06.078}{Nucl.\ Instrum.\ Meth.\  \textbf{A552} (2005) 566}, \href{http://arxiv.org/abs/physics/0503191}{{\normalfont\ttfamily arXiv:physics/0503191}}\relax
\mciteBstWouldAddEndPuncttrue
\mciteSetBstMidEndSepPunct{\mcitedefaultmidpunct}
{\mcitedefaultendpunct}{\mcitedefaultseppunct}\relax
\EndOfBibitem
\bibitem{PDG2024}
Particle Data Group, S.~Navas {\em et~al.}, \ifthenelse{\boolean{articletitles}}{\emph{{\href{http://pdg.lbl.gov/}{Review of particle physics}}}, }{}\href{https://doi.org/10.1103/PhysRevD.110.030001}{Phys.\ Rev.\  \textbf{D110} (2024) 030001}\relax
\mciteBstWouldAddEndPuncttrue
\mciteSetBstMidEndSepPunct{\mcitedefaultmidpunct}
{\mcitedefaultendpunct}{\mcitedefaultseppunct}\relax
\EndOfBibitem
\bibitem{AdaBoost}
Y.~Freund and R.~E. Schapire, \ifthenelse{\boolean{articletitles}}{\emph{A decision-theoretic generalization of on-line learning and an application to boosting}, }{}\href{https://doi.org/10.1006/jcss.1997.1504}{J.\ Comput.\ Syst.\ Sci.\  \textbf{55} (1997) 119}\relax
\mciteBstWouldAddEndPuncttrue
\mciteSetBstMidEndSepPunct{\mcitedefaultmidpunct}
{\mcitedefaultendpunct}{\mcitedefaultseppunct}\relax
\EndOfBibitem
\bibitem{Skwarnicki:1986xj}
T.~Skwarnicki, {\em {A study of the radiative cascade transitions between the Upsilon-prime and Upsilon resonances}}, PhD thesis, Institute of Nuclear Physics, Krakow, 1986, {\href{http://inspirehep.net/record/230779/}{DESY-F31-86-02}}\relax
\mciteBstWouldAddEndPuncttrue
\mciteSetBstMidEndSepPunct{\mcitedefaultmidpunct}
{\mcitedefaultendpunct}{\mcitedefaultseppunct}\relax
\EndOfBibitem
\bibitem{JohnsonSU}
N.~L. Johnson, \ifthenelse{\boolean{articletitles}}{\emph{{Systems of frequency curves generated by methods of translation}}, }{}\href{https://doi.org/10.1093/biomet/36.1-2.149}{Biometrika \textbf{36} (1949) 149}\relax
\mciteBstWouldAddEndPuncttrue
\mciteSetBstMidEndSepPunct{\mcitedefaultmidpunct}
{\mcitedefaultendpunct}{\mcitedefaultseppunct}\relax
\EndOfBibitem
\bibitem{Pivk:2004ty}
M.~Pivk and F.~R. Le~Diberder, \ifthenelse{\boolean{articletitles}}{\emph{{sPlot: A statistical tool to unfold data distributions}}, }{}\href{https://doi.org/10.1016/j.nima.2005.08.106}{Nucl.\ Instrum.\ Meth.\  \textbf{A555} (2005) 356}, \href{http://arxiv.org/abs/physics/0402083}{{\normalfont\ttfamily arXiv:physics/0402083}}\relax
\mciteBstWouldAddEndPuncttrue
\mciteSetBstMidEndSepPunct{\mcitedefaultmidpunct}
{\mcitedefaultendpunct}{\mcitedefaultseppunct}\relax
\EndOfBibitem
\bibitem{LHCb-DP-2013-002}
LHCb collaboration, R.~Aaij {\em et~al.}, \ifthenelse{\boolean{articletitles}}{\emph{{Measurement of the track reconstruction efficiency at LHCb}}, }{}\href{https://doi.org/10.1088/1748-0221/10/02/P02007}{JINST \textbf{10} (2015) P02007}, \href{http://arxiv.org/abs/1408.1251}{{\normalfont\ttfamily arXiv:1408.1251}}\relax
\mciteBstWouldAddEndPuncttrue
\mciteSetBstMidEndSepPunct{\mcitedefaultmidpunct}
{\mcitedefaultendpunct}{\mcitedefaultseppunct}\relax
\EndOfBibitem
\bibitem{LHCb-PUB-2016-021}
L.~Anderlini {\em et~al.}, \ifthenelse{\boolean{articletitles}}{\emph{{The PIDCalib package}}, }{} \href{http://cdsweb.cern.ch/search?p=LHCb-PUB-2016-021&f=reportnumber&action_search=Search&c=LHCb+Notes} {LHCb-PUB-2016-021}, 2016\relax
\mciteBstWouldAddEndPuncttrue
\mciteSetBstMidEndSepPunct{\mcitedefaultmidpunct}
{\mcitedefaultendpunct}{\mcitedefaultseppunct}\relax
\EndOfBibitem
\bibitem{CRANMER2001198}
K.~Cranmer, \ifthenelse{\boolean{articletitles}}{\emph{Kernel estimation in high-energy physics}, }{}\href{https://doi.org/https://doi.org/10.1016/S0010-4655(00)00243-5}{Comput.\ Phys.\ Commun.\  \textbf{136} (2001) 198}\relax
\mciteBstWouldAddEndPuncttrue
\mciteSetBstMidEndSepPunct{\mcitedefaultmidpunct}
{\mcitedefaultendpunct}{\mcitedefaultseppunct}\relax
\EndOfBibitem
\bibitem{Xie:2009rka}
Y.~Xie, \ifthenelse{\boolean{articletitles}}{\emph{{sFit: a method for background subtraction in maximum likelihood fit}}, }{}\href{http://arxiv.org/abs/0905.0724}{{\normalfont\ttfamily arXiv:0905.0724}}\relax
\mciteBstWouldAddEndPuncttrue
\mciteSetBstMidEndSepPunct{\mcitedefaultmidpunct}
{\mcitedefaultendpunct}{\mcitedefaultseppunct}\relax
\EndOfBibitem
\bibitem{Jackson:1964zd}
J.~D. Jackson, \ifthenelse{\boolean{articletitles}}{\emph{{Remarks on the phenomenological analysis of resonances}}, }{}\href{https://doi.org/10.1007/BF02750563}{Nuovo Cim.\  \textbf{34} (1964) 1644}\relax
\mciteBstWouldAddEndPuncttrue
\mciteSetBstMidEndSepPunct{\mcitedefaultmidpunct}
{\mcitedefaultendpunct}{\mcitedefaultseppunct}\relax
\EndOfBibitem
\bibitem{Narsky:1999kt}
I.~V. Narsky, \ifthenelse{\boolean{articletitles}}{\emph{{Estimation of upper limits using a Poisson statistic}}, }{}\href{https://doi.org/10.1016/S0168-9002(00)00305-3}{Nucl.\ Instrum.\ Meth.\  \textbf{A450} (2000) 444}, \href{http://arxiv.org/abs/hep-ex/9904025}{{\normalfont\ttfamily arXiv:hep-ex/9904025}}\relax
\mciteBstWouldAddEndPuncttrue
\mciteSetBstMidEndSepPunct{\mcitedefaultmidpunct}
{\mcitedefaultendpunct}{\mcitedefaultseppunct}\relax
\EndOfBibitem
\bibitem{LHCb-TDR-012}
LHCb collaboration, \ifthenelse{\boolean{articletitles}}{\emph{{Framework TDR for the LHCb Upgrade: Technical Design Report}}, }{} \href{http://cdsweb.cern.ch/search?p=CERN-LHCC-2012-007&f=reportnumber&action_search=Search&c=LHCb} {CERN-LHCC-2012-007}, 2012\relax
\mciteBstWouldAddEndPuncttrue
\mciteSetBstMidEndSepPunct{\mcitedefaultmidpunct}
{\mcitedefaultendpunct}{\mcitedefaultseppunct}\relax
\EndOfBibitem
\bibitem{LHCb-DP-2022-002}
LHCb collaboration, R.~Aaij {\em et~al.}, \ifthenelse{\boolean{articletitles}}{\emph{{The LHCb Upgrade I}}, }{}\href{https://doi.org/10.1088/1748-0221/19/05/P05065}{{JINST} \textbf{19} (2024) P05065}, \href{http://arxiv.org/abs/2305.10515}{{\normalfont\ttfamily arXiv:2305.10515}}\relax
\mciteBstWouldAddEndPuncttrue
\mciteSetBstMidEndSepPunct{\mcitedefaultmidpunct}
{\mcitedefaultendpunct}{\mcitedefaultseppunct}\relax
\EndOfBibitem
\bibitem{LHCb-PAPER-2022-002}
LHCb collaboration, R.~Aaij {\em et~al.}, \ifthenelse{\boolean{articletitles}}{\emph{{Amplitude analysis of the $\Lc \to \proton \Km\pip$ decay and $\Lc$ baryon polarization measurement in semileptonic beauty hadron decays}}, }{}\href{https://doi.org/10.1103/PhysRevD.108.012023}{Phys.\ Rev.\  \textbf{D108} (2023) 012023}, \href{http://arxiv.org/abs/2208.03262}{{\normalfont\ttfamily arXiv:2208.03262}}\relax
\mciteBstWouldAddEndPuncttrue
\mciteSetBstMidEndSepPunct{\mcitedefaultmidpunct}
{\mcitedefaultendpunct}{\mcitedefaultseppunct}\relax
\EndOfBibitem
\bibitem{LHCb-PAPER-2022-044}
LHCb collaboration, R.~Aaij {\em et~al.}, \ifthenelse{\boolean{articletitles}}{\emph{{$\Lc$ polarimetry using the dominant hadronic mode}}, }{}\href{https://doi.org/10.1007/JHEP07(2023)228}{JHEP \textbf{07} (2023) 228}, \href{http://arxiv.org/abs/2301.07010}{{\normalfont\ttfamily arXiv:2301.07010}}\relax
\mciteBstWouldAddEndPuncttrue
\mciteSetBstMidEndSepPunct{\mcitedefaultmidpunct}
{\mcitedefaultendpunct}{\mcitedefaultseppunct}\relax
\EndOfBibitem
\bibitem{LHCb-PAPER-2025-022-cds}
\url{http://cds.cern.ch/record/2938212}\relax
\mciteBstWouldAddEndPuncttrue
\mciteSetBstMidEndSepPunct{\mcitedefaultmidpunct}
{\mcitedefaultendpunct}{\mcitedefaultseppunct}\relax
\EndOfBibitem
\end{mcitethebibliography}
